\documentclass[times,twocolumn,final,authoryear]{elsarticle}
\usepackage{jasr}
\usepackage{framed,multirow}
\usepackage{amssymb}
\usepackage{latexsym}
\usepackage[switch]{lineno}
\usepackage{url}
\usepackage{xcolor}
\definecolor{newcolor}{rgb}{.8,.349,.1}
\usepackage{caption}
\usepackage{subcaption}

\usepackage{aas_journal_abrv}
\usepackage{hyperref}

\journal{Advances in Space Research}

\begin{document}

\verso{Richard Fallows \textit{et al.}}

\begin{frontmatter}

\title{Application of Novel Interplanetary Scintillation Visualisations using LOFAR: A Case Study of Merged CMEs from September 2017}

\author[1,5]{R.A. \snm{Fallows}\corref{cor1}}
\cortext[cor1]{Corresponding author: 
  email: rafallows@gmail.com
  }
\author[2]{K. \snm{Iwai}}
\author[3]{B.V. \snm{Jackson}}
\author[4,1]{P. \snm{Zhang}}
\author[5]{M.M. \snm{Bisi}}
\author[1]{P. \snm{Zucca}}

\address[1]{ASTRON - the Netherlands Institute for Radio Astronomy, Oude Hoogeveensedijk 4, 7991 PD Dwingeloo, the Netherlands}
\address[2]{Institute for Space-Earth Environmental Research, Nagoya University, Furo-cho, Chikusa-ku, Nagoya, 464-8601, Japan}
\address[3]{Center for Astrophysics and Space Sciences, University of California, San Diego, LaJolla, California, 92093-0424, USA}
\address[4]{Institute of Astronomy and National Astronomical Observatory,\\ Bulgarian Academy of Sciences, Sofia 1784, Bulgaria}
\address[5]{RAL Space, United Kingdom Research and Innovation – Science \& Technology Facilities Council – Rutherford Appleton Laboratory, Harwell Campus, Oxfordshire, OX11 0QX, UK}

\received{19 April 2022}
\finalform{22 August 2022}
\accepted{29 August 2022}
\availableonline{19 September 2022}
\communicated{S. Sarkar}

\begin{abstract}
Observations of interplanetary scintillation (IPS – the scintillation of compact radio sources due to density variations in the solar wind) enable the velocity of the solar wind to be determined, and its bulk density to be estimated, throughout the inner heliosphere.  A series of observations using the Low Frequency Array (LOFAR - a radio telescope centred on the Netherlands with stations across Europe) were undertaken using this technique to observe the passage of an ultra-fast CME which launched from the Sun following the X-class flare of 10 September 2017.  LOFAR observed the strong radio source 3C147 at an elongation of 82 degrees from the Sun over a period of more than 30 hours and observed a strong increase in speed to 900\,km\,s$^{-1}$ followed two hours later by a strong increase in the level of scintillation, interpreted as a strong increase in density.  Both speed and density remained enhanced for a period of more than seven hours, to beyond the period of observation.  Further analysis of these data demonstrates a view of magnetic-field rotation due to the passage of the CME, using advanced IPS techniques only available to a unique instrument such as LOFAR.
\end{abstract}


\end{frontmatter}


\section{Introduction}
\label{introduction}

Observations of interplanetary scintillation (IPS - \citet{Clarke:1964,Hewishetal:1964}) have been used for several decades to observe the solar wind and Coronal Mass Ejections (CMEs) throughout the inner heliosphere.  Such observations are typically used to estimate solar wind velocity \citep[e.g.][]{Coles:1996,ManoAnanth:1990,KojimaKakinuma:1990} and/or ``g-level'', a normalised measure of the strength of scintillation, related to density \citep[e.g.][]{Jacksonetal:1998,Tappin:1986}.  The IPS array operated by the Institute for Space-Earth Environmental Research (ISEE), Japan, in operation since the 1980s, is used to take regular measurements of these quantities \citep{Tokumaruetal:2011}.  These, in turn, are fed into a tomographic model to provide 3-D reconstructions of solar wind velocity and density throughout the inner heliosphere every six hours, and a five-day prediction (see \href{https://ips.ucsd.edu/high\_resolution\_predictions}{https://ips.ucsd.edu/high\_resolution\_predictions} - although heliospheric conditions resulting, e.g., the launch of a CME can be predicted at best only about two days ahead of time, co-rotating structures can be forecast at least five days ahead) of these values at the Sun-Earth L1 Lagrangian point \citep[e.g.][and references therein]{Jacksonetal:2020}.  Such views incorporate both the background solar wind and any CMEs present, so long as these were detected in the observations of IPS used as input.  Highly detailed views of the inner heliosphere are possible with this technique, given many more daily observations of sufficient quality than are regularly available at present \citep[][]{Bisietal:2009}.

As a transit instrument reliant upon Earth rotation to perform successive observations of the same set of radio sources, the ISEE IPS array is limited to only short-duration observations of each radio source every 24 hours, requiring the assistance of MHD simulations to reconstruct the fast propagating CMEs \citep[e.g.][]{Iwaietal:2021}.  Observing stations with radio source tracking capabilities  can make multiple passes during a single day, or dwell on individual sources for a longer period of time, and simultaneous observation from such system(s) with stations several hundred or more kilometres apart are capable of getting much more detail from single observations.  Such observations enable multiple solar wind streams to be detected crossing the observing station to radio source lines of sight, such as the ``fast and faster'' solar wind streams detected by the Ulysses spacecraft and observed in IPS using the combined European Incoherent Scatter (EISCAT) and Multi-Element Radio-Linked Interferometer Network (MERLIN) systems \citep[][]{Bisietal:2007}.  Furthermore, longer-duration observations are possible which enable changes in solar wind structure (e.g. due to the onset and/or the passage of a CME) to be tracked across a single line of sight.  For example, the onset of a CME from May 2005 was detected and part of its structure tracked in an observation taken using the EISCAT and MERLIN systems \citep[][]{Bisietal:2010,Changetal:2021}, and micro-scale structure in the slow solar wind observed in measurements by EISCAT \citep[e.g.][and references therein]{Hardwicketal:2013}.  It has also proved possible to detect an off-radial component to the fast solar wind, where \citet{Dorrianetal:2013} demonstrated that the polar solar wind shows a slight equatorwards expansion, and \citet{Breenetal:2008} noted that a fast stream adjacent to the May 2005 CME was deviated 8-15$^{\circ}$ off-radial by the CME itself.

\begin{figure}
    \centering
    \includegraphics[width=\linewidth]{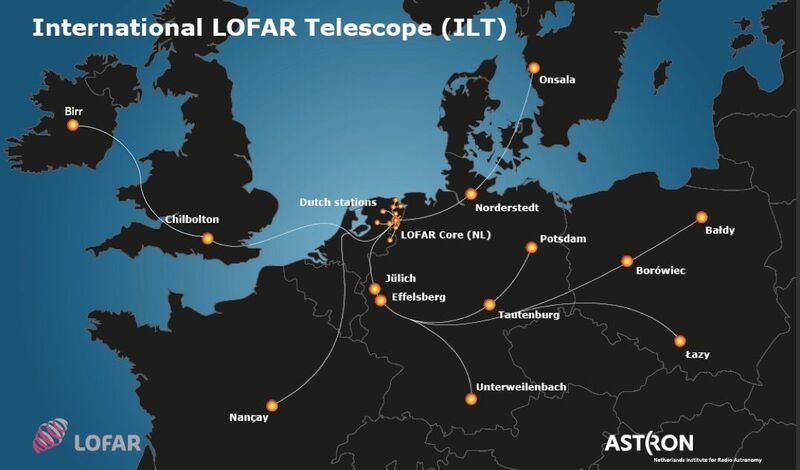}
    \caption{Map showing the distribution of LOFAR stations over Europe (6 in Germany, 3 in Poland, 1 each in France, Ireland, Sweden, and the UK, plus the Dutch array of 38 stations) at the time of observation in September 2017.  All stations are connected via dedicated high-speed data links to correlation and processing facilities in Groningen, Netherlands.}
    \label{fig:lofarmap}
\end{figure}

LOFAR (the low-frequency array, \citet{vanHaarlemetal:2013}) is Europe's largest and most flexible radio telescope, with capabilities which enable much more information to be extracted from multi-station observations of IPS.  The wide bandwidth enables any change in the scintillation pattern with frequency, e.g. between weak and strong scintillation, to be directly observed and features seen which would be invisible in a single-frequency measurement \citep[as observed in ionospheric scintillation measurements taken using LOFAR hardware, for example,][]{Fallowsetal:2014,Fallowsetal:2020}.  Furthermore, the international array contains 14 stations (at the time of writing - 13 were available at the time of the observations described here) outside the Netherlands with baselines of $\sim$200\,km to $>\sim$2000\,km, in addition to the Dutch array containing a dense ``core'' of stations and 14 ``remote'' stations scattered across the north-east of the Netherlands.   All stations are connected via dedicated high-speed data links to correlation and processing facilities in Groningen, Netherlands.  The array as it was in September 2017 is depicted in Figure \ref{fig:lofarmap} (it has since gained a new station near Ventspils in Latvia, a further station will be built near Medicina in Italy in 2023, and there are plans for a further station in Bulgaria).  This enables the spatial extent of the IPS correlation between stations to be investigated, leading to information on the density structure giving rise to the IPS to be studied, as will be detailed later in this paper.

September 2017 was the most active period of solar cycle 24, with three X-class flares, numerous M-class flares, and multiple CMEs.  The early arrival of the CME associated with the 6 September X-9 flare produced severe geomagnetic storming on 7 and 8 September; a further CME, thought at the time to be ultra-fast with a speed of $\sim$3000\,km\,s$^{-1}$, launched on 10 September associated with a further X-8.2 flare, as the active region responsible rotated around the solar limb.  A flurry of activity ensued following this latter event, as groups around the world attempted to find the CME in the heliosphere.  A Director's Discretionary Time (DDT) proposal was quickly submitted and approved, which enabled LOFAR to take observations of IPS for $\sim$30 hours from late morning on 11 September with the aim of finding the CME and tracking its passage across one or more lines of sight.

This paper details the observations taken, introduces analysis techniques which make full use of IPS data taken with LOFAR, and demonstrates the possibility for IPS to show magnetic field orientation as the CME passes across the line of sight.  Full MHD modelling of this event incorporating a comparison with the LOFAR results presented here is given in a companion paper by \citet{Iwaietal:2022}.

\section{LOFAR Observations}
\label{sec:observations}

At 15:35\,UT on 10 September 2017 an X8.2 flare was observed as active region AR12673 (then at S08, W88) rotated around the west limb of the Sun.  This was associated with an ultra-fast CME first observed in the LASCO C2 coronagraph at 16:00\,UT, with a velocity measured through the LASCO C3 field of view of 3,212\,km\,s$^{-1}$.  Full details of this event as seen by LASCO can be found via the Halo CME alert at \url{https://umbra.nascom.nasa.gov/lasco/observations/halo/2017/170910/}.  Later analyses, unavailable at the time, showed that this ultra-fast CME merged during its passage through the inner heliosphere with two slow CMEs which had launched on 9 September 2017 and themselves merged whilst still within the  LASCO C3 field of view \citep[e.g.][]{Guoetal:2018,Leeetal:2018}.  The LOFAR observations detailed here therefore observed the result of the merger of all three CMEs, rather than the single event originally envisaged, and at a later time than expected.

\begin{figure}
    \centering
    \includegraphics[trim=180 150 0 140,clip,width=\linewidth]{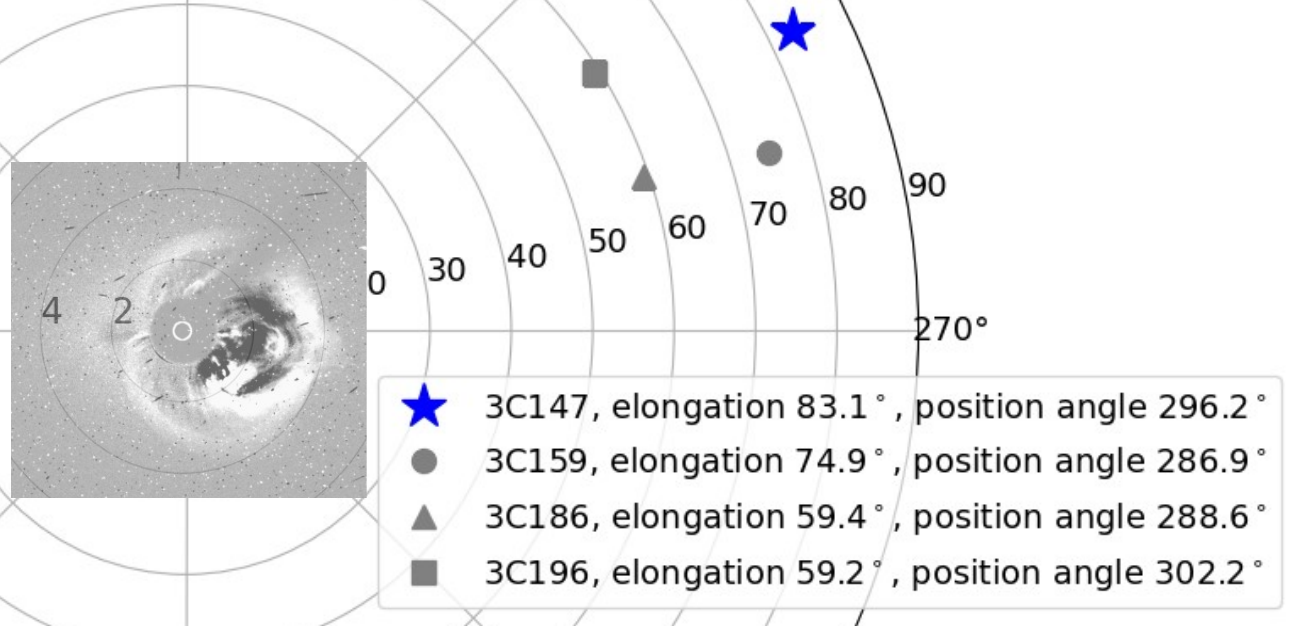}
    \caption{Segment of a polar plot giving the elongations and position angles (clock angle anti-clockwise from solar north) in the sky plane of the radio sources observed by LOFAR.  A LASCO C3 difference image taken at 16:54\,UT on 10 September 2017 is superposed for reference (not displayed to scale).}
    \label{fig:polarlascosources}
\end{figure}

LOFAR observations were carried out from 11:30\,UT on 11 September to 14:00\,UT on 12 September 2017 to observe this event, and alternated between four radio sources, chosen such that it was considered likely that the CME would pass across the line of sight to at least one of them.  Figure \ref{fig:polarlascosources} gives a plot of the locations of the sources used, along with a reference LASCO C3 difference image from the time.  Observations were of 9\,minutes duration, with an obligatory 1\,minute gap in between, alternating between sources during the periods when any source was above an elevation of 25$^{\circ}$ as seen from the LOFAR core.  The exception to this was 3C147 which is circumpolar from LOFAR latitudes and for which observations continued throughout.  The observing scheme is detailed in Table \ref{tab:observations}.

All observations recorded Stokes-I dynamic spectra for 400 subbands, each 195.3125\,kHz wide, covering contiguously the frequency range 110-190\,MHz, with an integration time of 0.01\,s.  Dynamic spectra were recorded individually for each international and Dutch remote station included in the observation (only the station near Potsdam, Germany, was unavailable due to a fault at the time) and for a single tied-array beam from the coherently-combined core stations (which is assumed in subsequent analysis as being equivalent to a single station located at the centre of the core, with the coordinates of station CS002LBA). 

\begin{table}
    \centering
    \begin{tabular}{|c|c|l|}
         \hline
         {\bf Date} & {\bf Period} & {\bf Sources Observed} \\
         \hline
         2017-09-11 & 11:30 - 14:29\,UT & 3C186,3C159,3C147 \\
         2017-09-11 & 14:30 - 16:19\,UT & 3C196,3C147 \\
         2017-09-11 & 16:20 - 23:59\,UT & 3C147 \\
         2017-09-12 & 00:00 - 13:59\,UT & 3C159,3C147 \\
         \hline
    \end{tabular}
    \caption{Periods of observation of each radio source.  All observations were of 9\,minutes duration with a 1\,minute gap between.  The sources are noted in the order of observation so, for example, on 11 September 3C186 was observed 11:30-11:39\,UT, 3C159 11:40-11:49\,UT, 3C147 11:50-11:59\,UT, back to 3C186 etc.}
    \label{tab:observations}
\end{table}

\begin{figure}
    \centering
    \includegraphics[width=\linewidth]{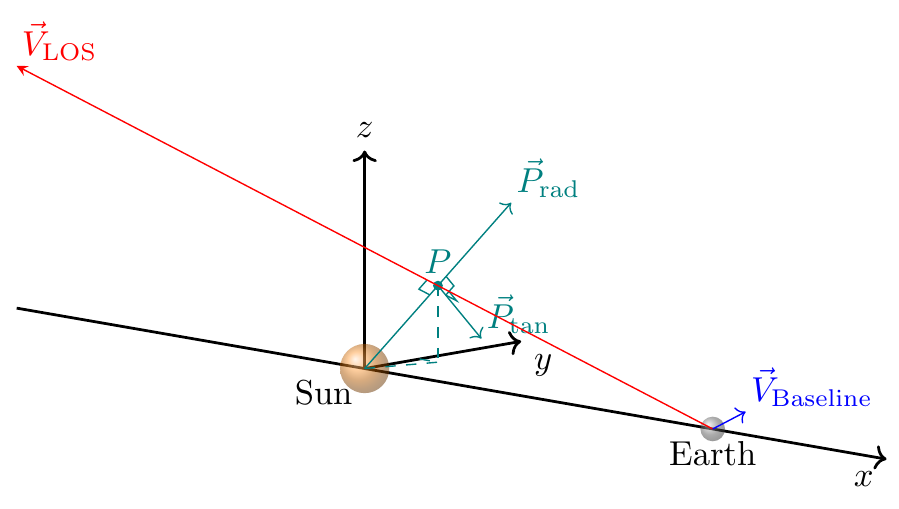}
    \caption{Schematic plot of coordinates for an observation of IPS.  The red vector marks the line of sight, the blue vector marks the direction of a baseline between a pair of observing stations, and the green vectors mark the components of the projection of this baseline onto a plane perpendicular to the line of sight at the P-point - the point of closest approach of the line of sight to the Sun.}
    \label{fig:coord}
\end{figure}

Before detailing the initial processing of these data it is helpful to describe the coordinate system used with reference to the line of sight between the radio source and an observing station on Earth, as illustrated by the schematic in Figure \ref{fig:coord}.  Although the line of sight cuts through an extended portion of the inner heliosphere, the majority of scattering typically comes from around the point of closest approach (the so-called ``P-point'') of the line of sight to the Sun (as detailed in Section \ref{sec:analysis}).  Therefore, the location of the P-point is usually used to define the coordinates of the observation relative to the Sun (e.g., a heliographic latitude and distance from the Sun).  

When an observation uses more than one station, the physical baseline between each pair of stations can be projected onto the sky-plane in the direction of the radio source from the Earth and expressed in terms of components in the radial direction from the Sun and tangential to it (green vectors in Figure \ref{fig:coord}).  The values of these components naturally change during the course of an observation as the Earth rotates (see descriptions of the geometry given in \citet{Dorrianetal:2013} and \citet{Moranetal:1998}), so mean values are used in the analysis of each observation. 

\begin{figure}
    \centering
    \includegraphics[width=\linewidth]{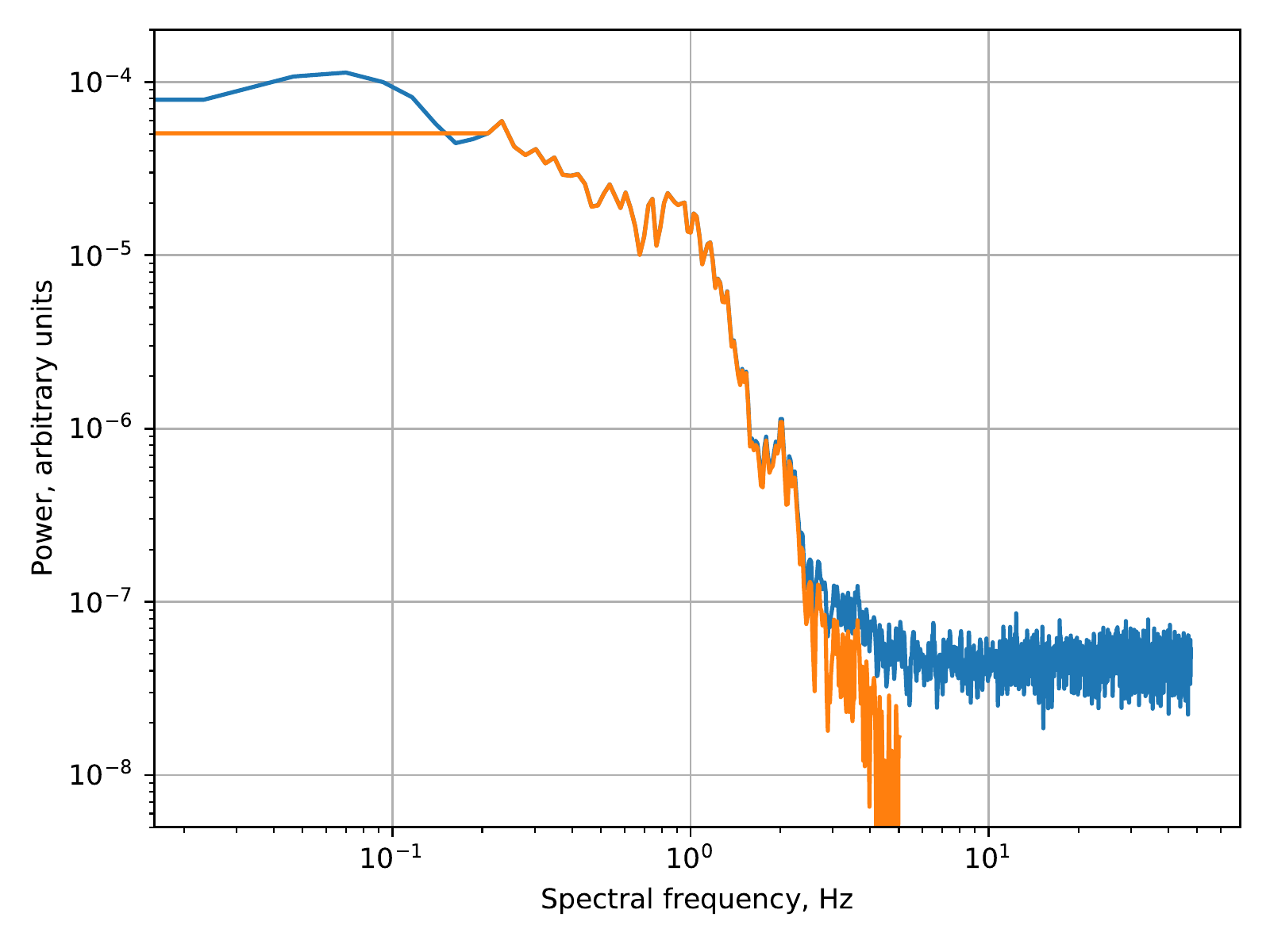}
    \caption{Example raw power spectrum and spectrum after filtering and noise subtraction, calculated using SE607 data from the 9-minute observation of 3C147 taken at 05:30\,UT on 12 September 2017.}
    \label{fig:spectra}
\end{figure}

Initial processing consisted of mitigating the effects of radio-frequency interference (RFI), time series calculation, and calculation of auto- and cross-correlation functions from each 9-minute observation.  This was carried out following a similar procedure to that of \citet{Fallowsetal:2020} for ionospheric scintillation, but with some parameter differences:
\begin{itemize}
    \item RFI mitigation: A median filter was applied to the dynamic spectra using a window of (1.95\,MHz $\times$ 0.525\,s) and then the original data divided by the median-filtered version to flatten out the scintillation pattern.  RFI were then identified as absolute values greater than 10\,$\sigma$, where $\sigma$ is the median absolute deviation (MAD) of the flattened dataset.  The MAD is used because the RFI can manifest as extreme outliers in the data, making this measure more robust than the standard deviation.  Data points identified as RFI are flagged and not used in further processing.
    \item Time series' of intensity received by each station are calculated by averaging the intensities for each time sample over the full frequency band of 110--190\,MHz; this is reasonable since the scintillation pattern remains highly correlated over the band in this set of observations.
    \item Calculate auto- power spectra using the intensity time series' from each station.
    \item Apply a high-pass filter at 0.2\,Hz to exclude the DC-component and any obvious ionospheric scintillation or slow system variation at the low spectral frequencies, and a low-pass filter ($f_c$) at 5\,Hz to cut out white noise at the high spectral frequencies.  The white noise is also subtracted using an average of spectral power over the high frequencies above the low-pass filter value.  An example is shown in Figure \ref{fig:spectra}.
    \item  Calculate auto-correlation functions using the filtered power spectra.
    \item Cross- power spectra and cross-correlation functions were calculated for time series' from every pair of stations (a total of 351 combinations) following the same methods.
    \item The baseline between every pair of stations was projected onto the sky-plane and components in the radial direction from the Sun and tangential to it calculated.
\end{itemize}

The processing described above can only mitigate the effects of short-duration and/or narrow-band spikes of RFI, so a further selection process is necessary to try and exclude stations more strongly affected by interference in any given observations, or suffering from weak signal-to-noise for other reasons (e.g., the remote stations contain half the number of high-band antennas -  high-band refers to the frequency range, which covers that used here - compared to the international stations, and the station in Ireland was newly-built and not yet well-calibrated at the time of observation).  These effects are most obvious in the auto-correlation functions which should be more or less the same for simultaneous data taken from different stations.  The auto-correlation functions of unreliable data tend to exhibit a much narrower peak than those of reliable data.  Hence datasets whose normalised auto-correlation function values close to the peak deviated by more than one standard deviation from the median for all auto-correlation functions in the observation were excluded from further analysis.

The cross-correlation function typically exhibits one or more peaks at one or more time-lags, corresponding to the velocity(s) of material crossing the lines of sight and so dependent on the length of the projected baseline between the pair of stations correlated.  Since this is only sensitive to material crossing perpendicularly to the lines of sight, velocity estimates presented here are in the sky plane and thus represent foreshortened versions of the true values for locations along the line of sight away from the P-point. 


\begin{figure}
    \centering
    \includegraphics[width=\linewidth]{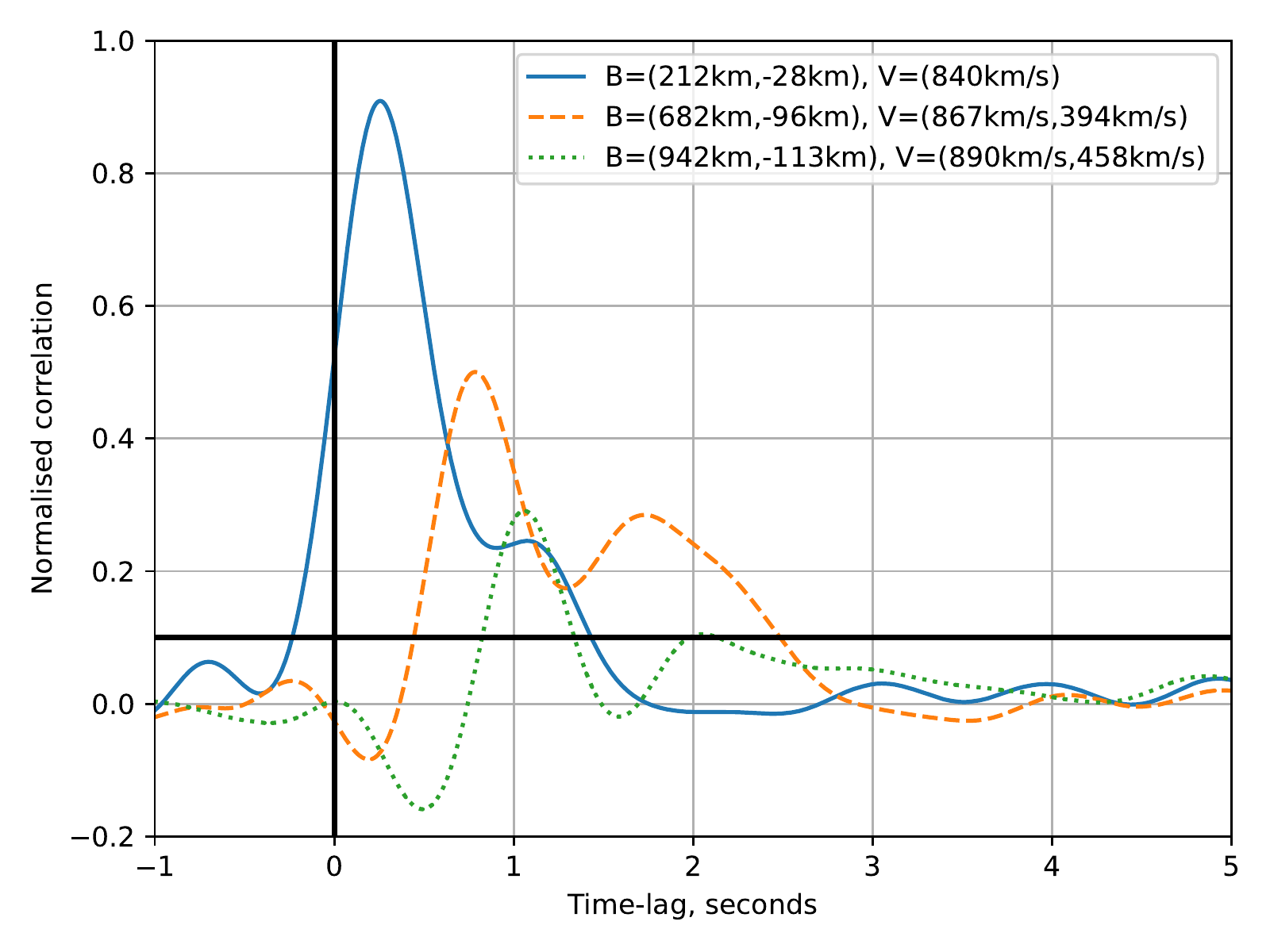}\\ \includegraphics[width=\linewidth]{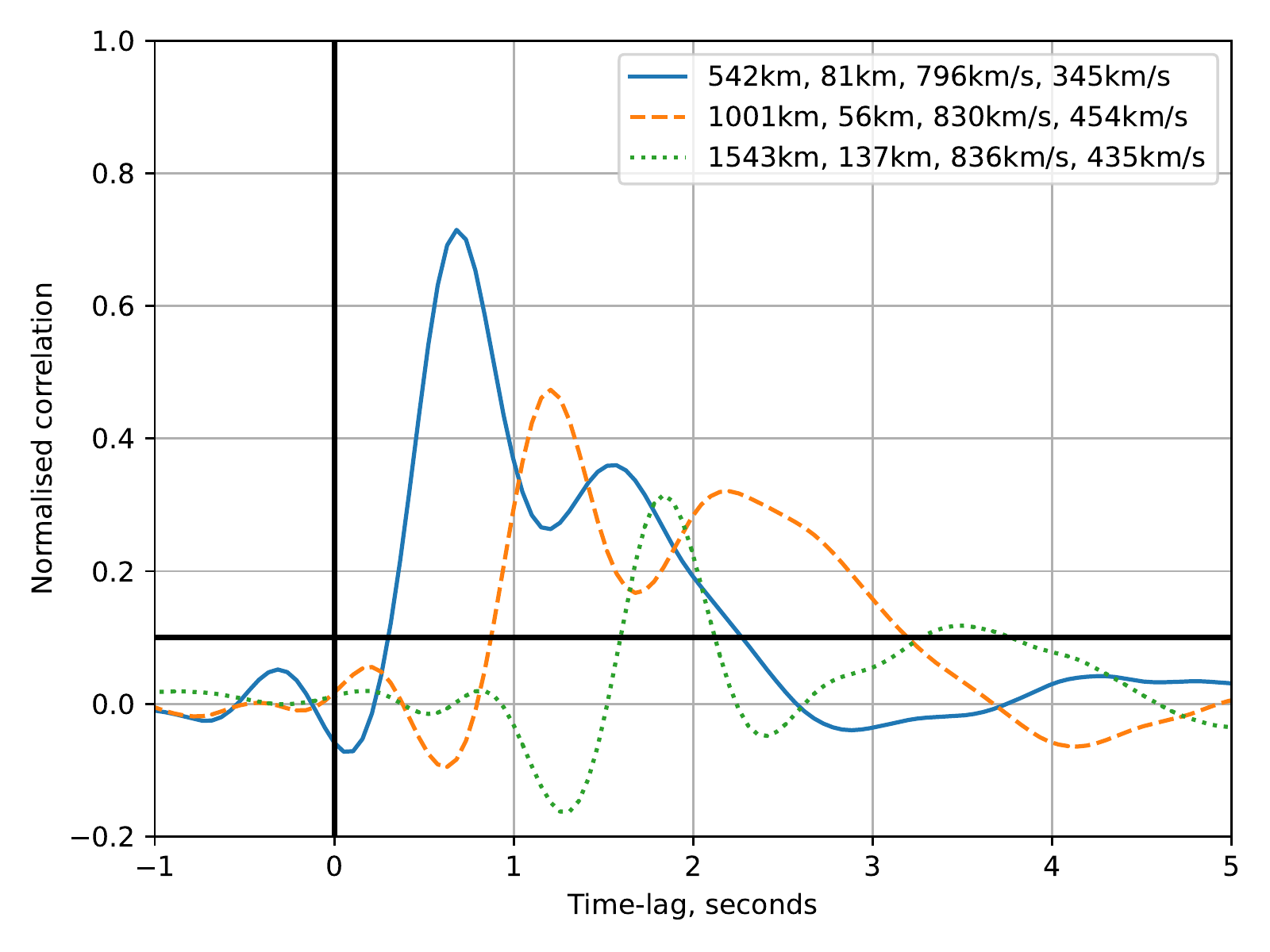}
    \caption{Two sets of cross-correlation functions (CCFs) from the 9\,minute observation of 3C147 taken at 05:30\,UT on 12 September 2017.  Each set shows three functions covering a range of baseline lengths along a roughly-consistent direction, a few degrees away from the radial direction.  The figures quoted in the legends are baselines B, (radial baseline, tangential baseline), a simple velocity calculation, V, of radial baseline divided by the time lag of the main peak, and the equivalent velocity calculation for a second peak where this exists.  The thick black horizontal line denotes the level above which correlation is deemed significant.  Top: Baseline direction tending towards the ecliptic plane. Cross-correlations are between stations DE609-RS509, DE602-FR606, and PL612-RS409 for the shortest to longest baselines respectively.  Bottom: Baseline direction tending north, away from the ecliptic plane. Cross-correlations are between stations DE605-UK608, PL611-DE605, and PL611-UK608 for the shortest to longest baselines respectively.}
    \label{fig:ccfs}
\end{figure}

A set of example cross-correlation functions (CCFs) from the observations presented here is given in Figure \ref{fig:ccfs}, which illustrate the effect of increasing baseline length \citep[see also the model CCFs given in][]{Coles:1996}.  Two different solar wind velocities were detected in this observation; a regular slow solar wind stream and that of the faster CME. The CCF from the pair of stations with the shortest baseline displayed here (212\,km radial, blue solid curve in Figure \ref{fig:ccfs} top) registers the second, slow, stream only as a bump at a time lag of just over 1\,s, but this bump becomes a second distinct peak in the CCFs as the baseline length increases, allowing a more direct measurement of the velocity it corresponds to. 

Since very few almost purely radial baselines are available, Figure \ref{fig:ccfs} shows two sets of baselines, each following very approximately a different off-radial direction (no baselines are exactly radial in the dataset), one tending a few degrees polewards (bottom) and the other equatorwards (top).  This illustrates a further interesting aspect: In the lower set of plots, the slow stream part of the cross-correlation functions (covering time-lags 2-3\,s) appears to broaden into a further ``bump'', suggesting the presence of a third, slightly slower, stream.  This is less apparent in the upper set of plots, possibly indicating that this third stream is slightly off-radial in its direction.

\section{Analysis}
\label{sec:analysis}

IPS is the result of an integral of scattering taking place along an extended line of sight.  Typically, scattering is assumed to be ``weak'', an assumption which is valid into an approximate elongation from the Sun which is dependent on observing frequency \citep[][]{Coles:1978} and solar wind conditions (e.g., the weak scattering assumption remains valid closer to the Sun for the  less-dense fast solar wind above polar coronal holes than the denser slow wind, as demonstrated in, e.g., \citet{Manoharan:1993} and \citet{Fallowsetal:2002}).  Closer to the Sun this assumption breaks down, although it has been demonstrated to remain valid for cross-correlation analyses well into the transition between weak and strong \citep[e.g.][]{Kojimaetal:2013}.  For the frequency range 110-190\,MHz the transition between weak and strong is expected to occur over elongations of $\sim13-25^{\circ}$.  However, the transition is obvious in LOFAR dynamic spectra, as demonstrated for ionospheric scintillation by \citet{Fallowsetal:2014}.

The IPS temporal power spectrum under weak scattering conditions can be 
expressed by the following equations, using the formulations given in \citet{Scottetal:1983} and \citet{Yamauchietal:1996}, separated in similar fashion to the latter reference for ease of description: 

\begin{eqnarray}
    \label{eqn:pspec}
    P(f) = C \lambda^2 \int_0^\infty \frac{2\pi}{v_{p}(z)}
    \int_{-\infty}^\infty F_{diff}({\bf q},z) F_{source}({\bf q},z) \nonumber \\ 
    \Phi_{ne}({\bf q},z) dq_y dz
\end{eqnarray}

\noindent Here, $\lambda$ is the observing wavelength; $C$ is a constant of proportionality, and $v_p$ is the component of solar wind velocity perpendicular to the line of sight.  ${\bf q}$ = $(q_x,q_y,q_z)$ is the three-dimensional spatial wavenumber vector for a co-ordinate system where $z$ is along the line of sight away from Earth, and $x$ and $y$ are perpendicular to the line of sight with $x$ being radial in direction from the Sun as viewed in the sky plane.

The Fresnel propagation filter acts as a high-pass filter blocking wavenumbers ${\bf q}$ below the Fresnel frequency $q_{f} = \sqrt{\frac{4 \pi}{\lambda z}}$:

\begin{equation}
    \label{eqn:Fresnel}
    F_{diff}({\bf q},z) = sin^2 \left( \frac{{\bf q}^2 \lambda z}{4 \pi} \right)
\end{equation}

The radio source observed is rarely the point source often assumed and has structure which acts as a low-pass filter and can be described by a source visibility function:

\begin{equation}
    \label{eqn:fsource}
    F_{source}({\bf q},z) = |V({\bf q},z)|^2
\end{equation}

The spatial spectrum of the density variations causing the observed scintillation is described as:

\begin{equation}
    \label{eqn:phine}
    \Phi_{ne}({\bf q},z) \propto {\bf q}^{-\alpha} exp \left(-{\left( \frac{\bf q}{q_i} \right)}^2 \right) r^{-4}
\end{equation}

\noindent where: $q_i$ is the wavenumber of an ``inner scale'' which is the dissipation scale of the density fluctuations; $\alpha$ is the power law index of the turbulent density spectrum; and $r$ is distance from the Sun to the scattering screen.  The density fluctuations are typically assumed to be anisotropic and elongated along the magnetic field, itself usually assumed to be radial in direction at the distances at which most observations of IPS are taken.  Since it is only structure in the $(x,y)$ plane which affects the scattering, the following definition is used: 

\begin{equation}
    \label{eqn:q}
    q = \sqrt{ q_x^2 + \left( \frac{q_y}{AR} \right)^2 }
\end{equation}

\noindent where AR is the axial ratio defining elongation of the fluctuations.  Since scattering falls as $r^{-4}$ (Equation \ref{eqn:phine}), it is mostly assumed to be coming from around the point of closest approach of the line of sight to the Sun.  However, dense material crossing elsewhere can make a significant contribution, as demonstrated by the existence of multiple peaks in the cross-correlation functions of Figure \ref{fig:ccfs} \citep[and illustrated in, e.g.,][]{Breenetal:2008}.  Therefore integration along the line of sight is normally performed under the assumption that scattering takes place due to a number of ``thin screens'' along the line of sight,  which each have their own parameters and are summed using the first-order Born approximation \citep[e.g.][]{ColesHarmon:1978, Coles:1996}.

This formed the basis of the weak scattering model used to analyse observations of IPS using EISCAT and MERLIN, as described in \citet{Fallowsetal:2008}.  However, these codes could not be easily ported to new hardware and adapted for use with LOFAR data, so to date only basic velocity calculations (a simple average of radial velocities calculated using the time-lags of each cross-correlation function) have been made which were then modelled using tomography.  The tomographic modelling process deals with biases due to integration along an extended line of sight and performs fitting which mitigates for errors in individual observations.  These results have been found to be consistent with those produced using ISEE data alone, and the tomographic modelling improved by the inclusion of the additional LOFAR data \citep[e.g.][and references therein]{Jacksonetal:2020}.

The more detailed analyses and techniques described in subsequent subsections illustrate the wealth of information which can be obtained from the many baselines available with LOFAR, even without applying a full weak-scattering model fit to the data.  However, the description given here is necessary to provide the full context and understanding of the techniques described.

\subsection{Velocity}
\label{subsec:velocity}

\begin{figure}
    \centering
    \begin{subfigure}{\linewidth}
        \centering
        \includegraphics[width=\linewidth]{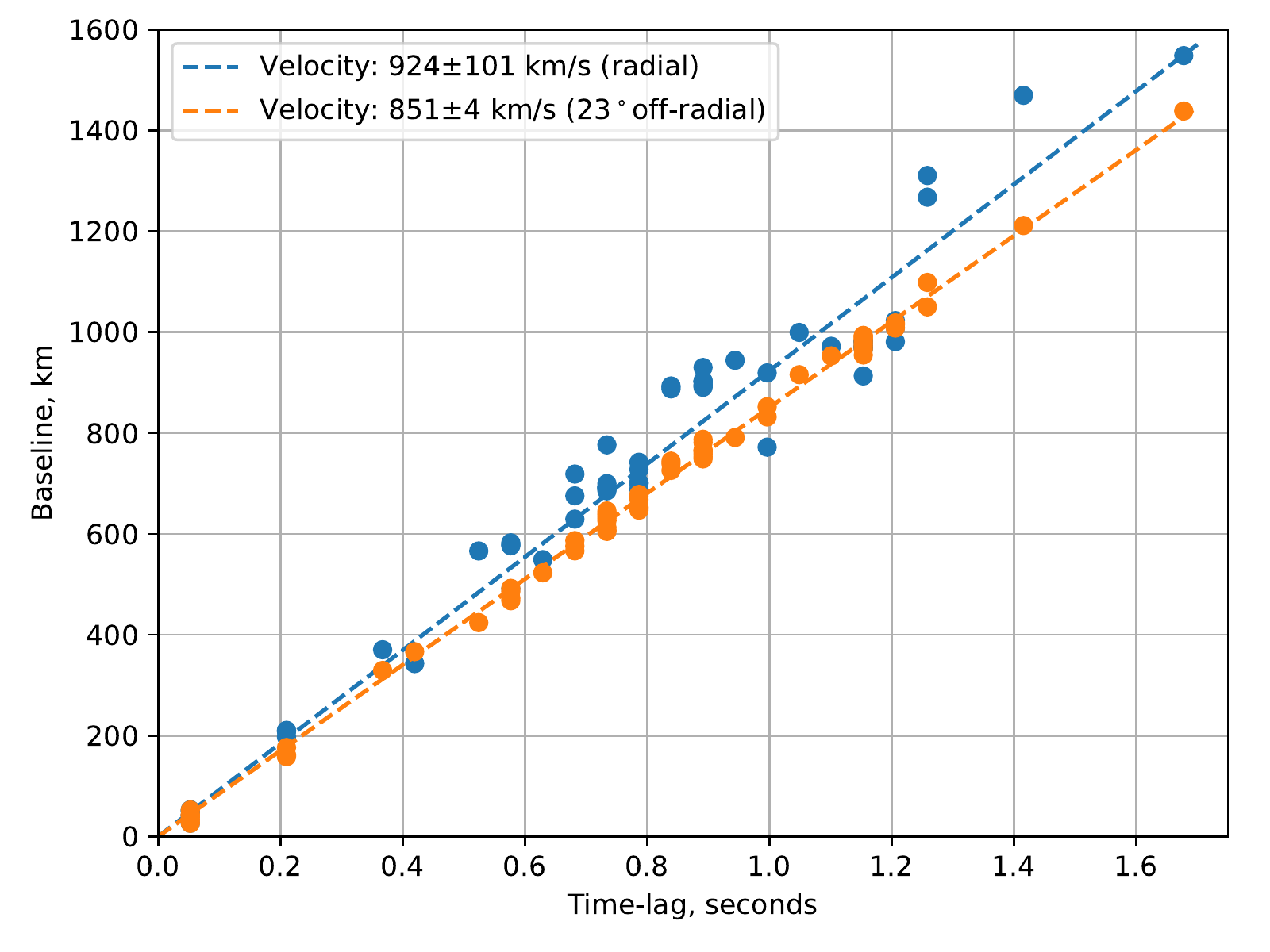}
        \caption{}
        \label{subfig:velocityfits}
    \end{subfigure}
    \begin{subfigure}{\linewidth}
        \centering
        \includegraphics[width=\linewidth]{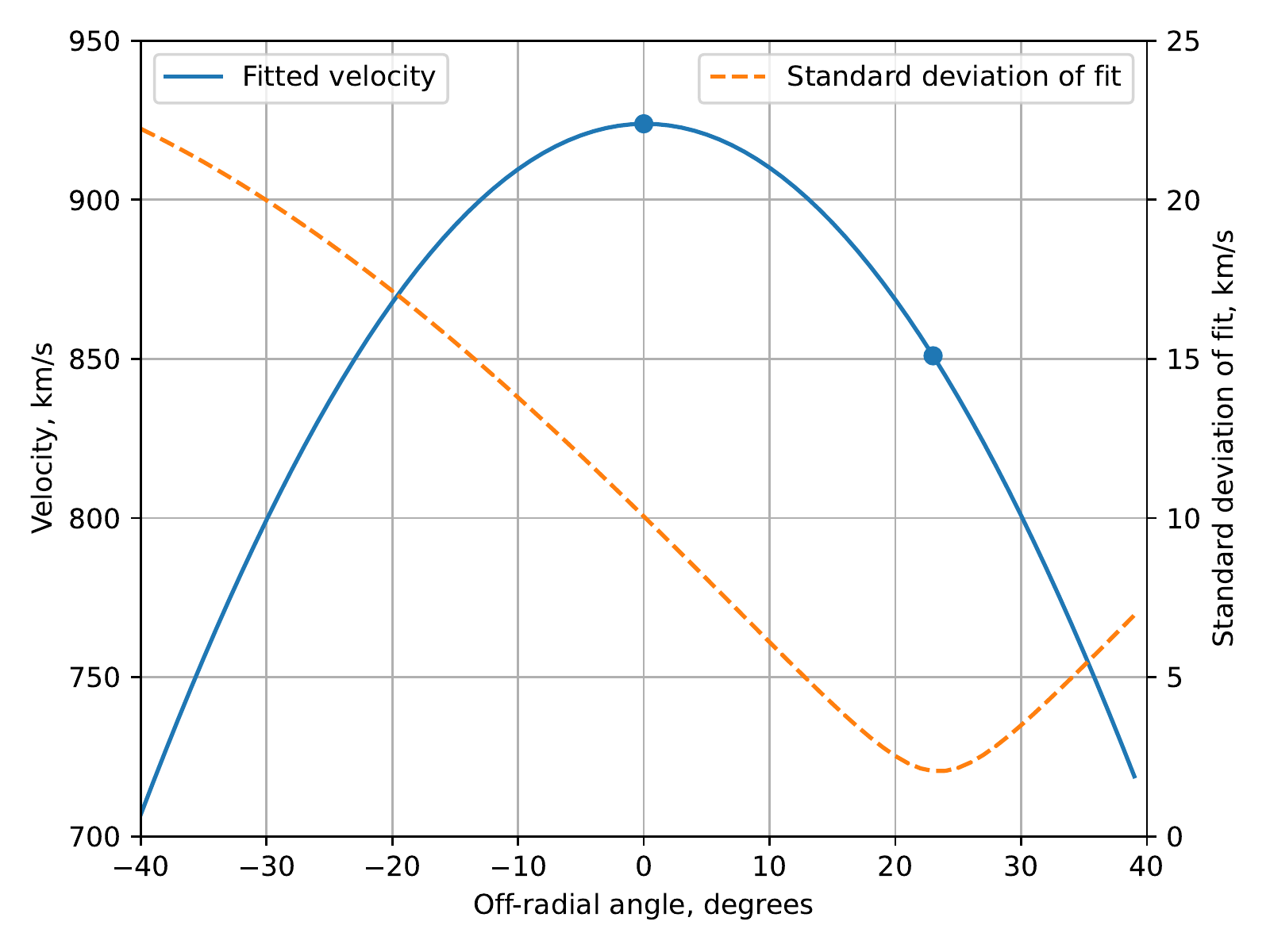}
        \caption{}
        \label{subfig:anglesds}
    \end{subfigure}
    \caption{(a) Fits to baseline versus time lag of maximum cross-correlation for the components of baselines in the radial direction (blue) and 23 degrees off-radial (towards the solar pole, orange), for the observation of 3C147 taken at 05:50\,UT on 12 September 2017.  The two points on the plot indicate the values of the fits for the angle of maximum velocity and that of the minimum standard deviation. (b) Plots of fitted velocity (blue) and standard deviation of the fit (orange) for fits assuming a range of off-radial angles.  Angles towards the solar pole are positive, towards the equator negative.} 
    \label{fig:velocity0550}
\end{figure}

Typically, the solar wind is assumed to be radial in direction.  Velocity estimates can be found by performing a least-squares fit to a plot of radial baseline versus the time-lags of the peak(s) of the CCFs calculated for a given observation or time segment thereof.  This is illustrated for the observation of 3C147 taken at 05:50\,UT on 12 September 2017 in Figure \ref{subfig:velocityfits}.

A degree of scatter is seen, which is reduced considerably if an off-radial direction of the velocity is assumed (Figure \ref{subfig:velocityfits}).  Optimum directions to assume for the velocity can be calculated following similar methods to those given in \citet{Fallowsetal:2020}, by fitting velocity and standard deviation for a range of off-radial angles, as given in Figure \ref{subfig:anglesds}.  However, whereas \citet{Fallowsetal:2020} found the direction of velocity for an ionospheric scintillation measurement to be equal for both a maximised velocity and minimised standard deviation, these two directions differ considerably here.  The direction of maximised velocity is basically radial in this instance, whereas the direction of minimum standard deviation is heavily biased by the orientation of the spatial density pattern, itself incorporating a bias due to a slightly elongated radio source structure, as detailed in Section \ref{subsec:spatialcorrelation}.  Hence the direction of maximised velocity is considered to be broadly consistent with the solar wind flow direction in subsequent analyses.

\begin{figure}
    \centering
    \includegraphics[width=\linewidth]{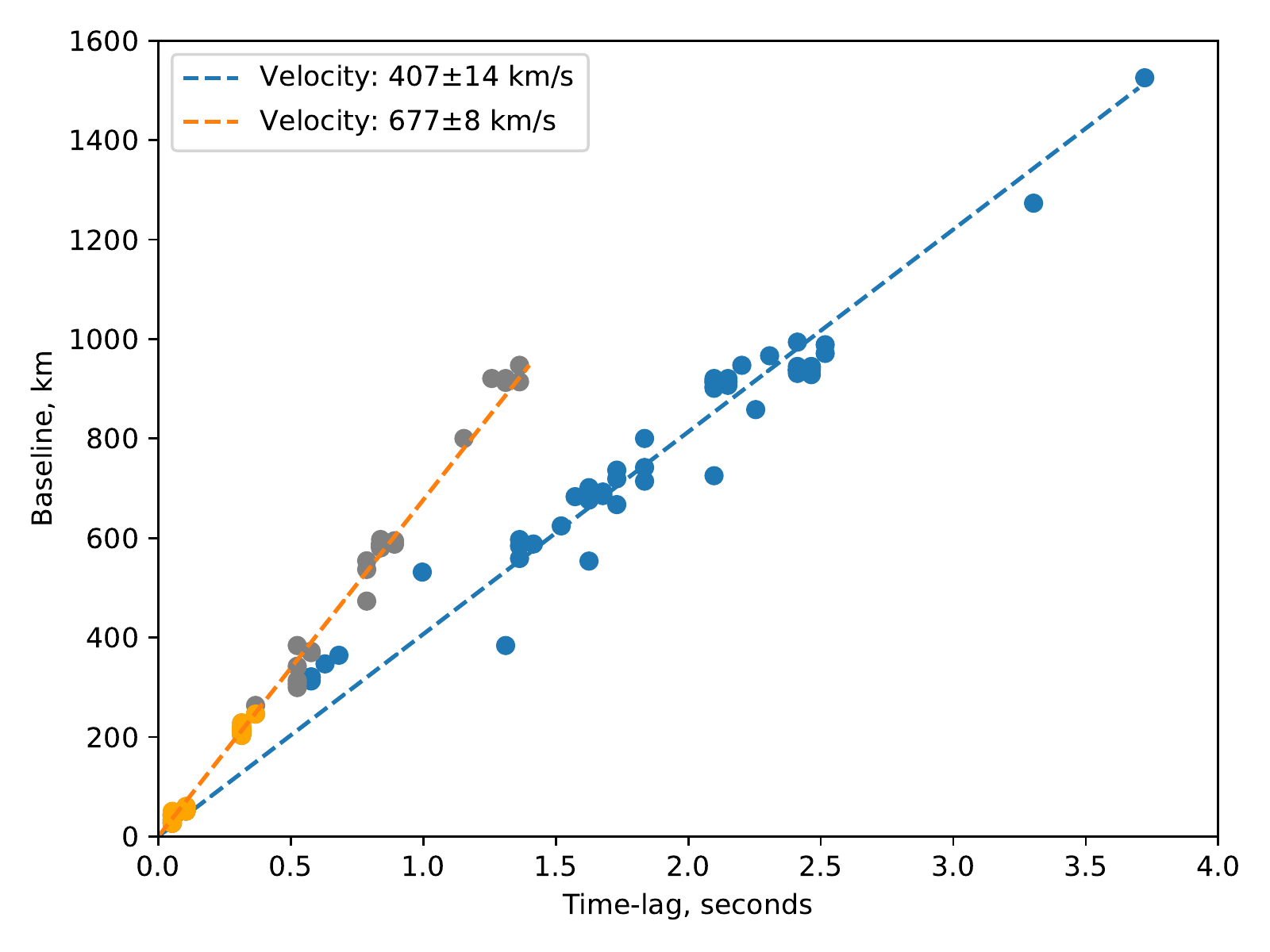}
    \caption{Fits to radial baseline versus time lag of cross-correlation peaks, separated according to the velocities indicated, for the observation of 3C147 taken at 05:10\,UT on 12 September 2017.  Points unused in either fit are indicated in grey.}
    \label{fig:twovelocity}
\end{figure}

The presence of two distinct velocities is also clear on these baseline versus time-lag plots.  Figure \ref{fig:twovelocity} shows an illustration of this, where two clear tracks of points are seen, a main track corresponding to a velocity consistent with a slow solar wind stream (blue points), and a secondary track indicating a faster velocity associated with CME material (orange and grey points).  The points and line fits displayed here are for baselines in the radial direction, for illustration.  Reliable determination of which points best correspond to which track in a case like this can be challenging.  In general, points with the longest baselines correspond to the track with the lower velocity, and those with the shortest baselines to the track with the higher velocity.  However, in some cases in this set of observations the high-velocity track can extend to longer baselines.  Therefore the points to associate with the main track (whether that is fast or slow) were determined by taking points with radial velocities within $\pm$100\,km\,s$^{-1}$ of the velocity corresponding to the main peak in a histogram of radial velocities calculated for all points.  Any ``second'' velocity was determined from a fit to all data points with baselines under 250\,km.  While this method can doubtless be improved, it proved to be reliable for all observations.  In the example given in Figure \ref{fig:twovelocity}, data points unused in either fit are plotted in grey.

\begin{figure}
    \centering
    \includegraphics[width=\linewidth]{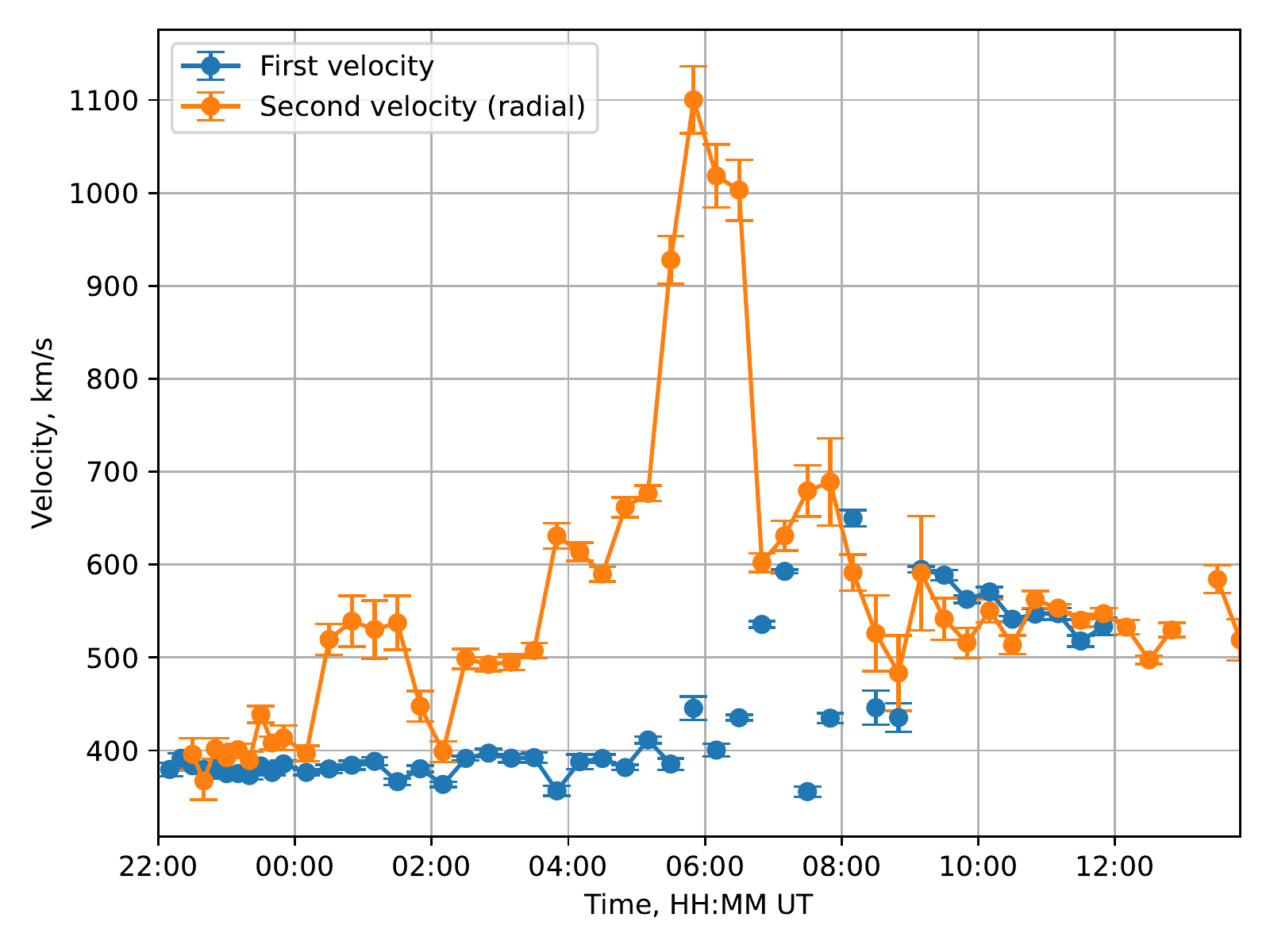}
    \includegraphics[width=\linewidth]{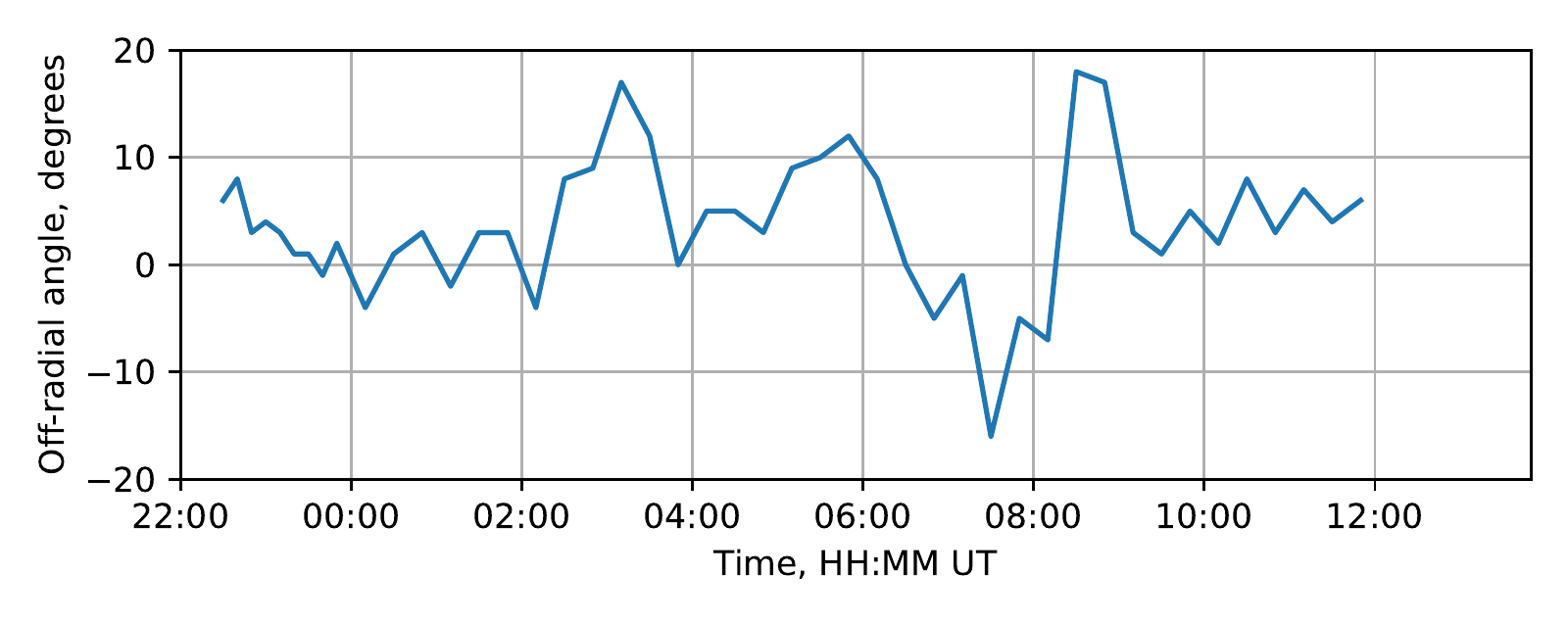}
    \caption{Top: Velocities calculated from observations taken from 22:00\,UT on 11 September to 12:00\,UT on 12 September 2017 for the flow direction of maximised velocity from fits to all baselines ("First velocity" - blue), and only to radial baselines under 250\,km ("Second velocity" - orange).  Error bars are the estimated standard deviation resulting from the fits.  Bottom: Deviations from the radial direction of the maximised fitted velocities, for the blue "first" velocity points.  Points after 12:00\,UT are considered unreliable and so not shown. }
    \label{fig:velocities}
\end{figure}

These techniques were applied to the observations of 3C147 taken between 22:00\,UT on 11 September and 13:50\,UT on 12 September 2017, as shown in Figure \ref{fig:velocities}. The observations of 3C147 taken prior to this period suffered from generally poor signal in the received scintillation intensities and therefore poor-quality CCFs (as, unfortunately, did the observations of other sources).  This indicates that the scintillation is generally too weak for good-quality observation by LOFAR at these source elongations from the Sun, even for a relatively strong radio source such as 3C147.  The time at which good-quality scintillation signals appear may in itself then indicate the time from which the effects of the CME start to manifest themselves in the received scintillation signal, even though the velocities at this time appear to be dominated by a regular slow solar wind stream.  After 12:00\,UT the data quality degrades significantly and so only the radial velocities calculated using short baselines are shown.  The off-radial angles displayed are those assuming a flow direction equivalent to that of the maximised velocity in the fitting process.

The results clearly show the impact of CME material crossing the lines of sight, with a general rise in velocity from shortly after 00:00\,UT to a sharp peak, clearly distinct from a background, slow, solar wind component.  Drops in the second (CME-related) velocity to the level of the slow solar wind at around 02:00\,UT and 08:30\,UT may indicate that the CME becomes briefly indistinct in the received scintillation.  From around 09:00\,UT the slow solar wind component is indistinct, with the scintillation pattern then dominated by the declining velocity of the CME material.

The direction of the maximised velocities corresponding to the slow background solar wind (blue curve in the lower plot of Figure \ref{fig:velocities} appears to be predominantly radial in the lead-up to the main increase in fast velocity (although some deviation is seen at around 23:00\,UT, ahead of any rise in velocity), but substantially deviates to nearly 15$^\circ$ off-radial in the polar direction around 03:00\,UT and remaining reasonably variable afterwards, over-shooting in an equatorial direction at around 07:00\,UT, and then tending more back to the radial direction.  Directions associated with the CME velocities exhibited large swings between $\pm30^\circ$, but these velocities are primarily visible, and fitted, to short-baseline data which show a higher degree of scatter and there is little evidence for such off-radial components in the analyses presented in sub-section \ref{subsec:spatialcorrelation}.  Hence these non-radial determinations are not shown, and the velocities plotted in orange in Figure \ref{fig:velocities} are those calculated for the radial direction only.

\begin{figure}
    \centering
    \includegraphics[width=\linewidth]{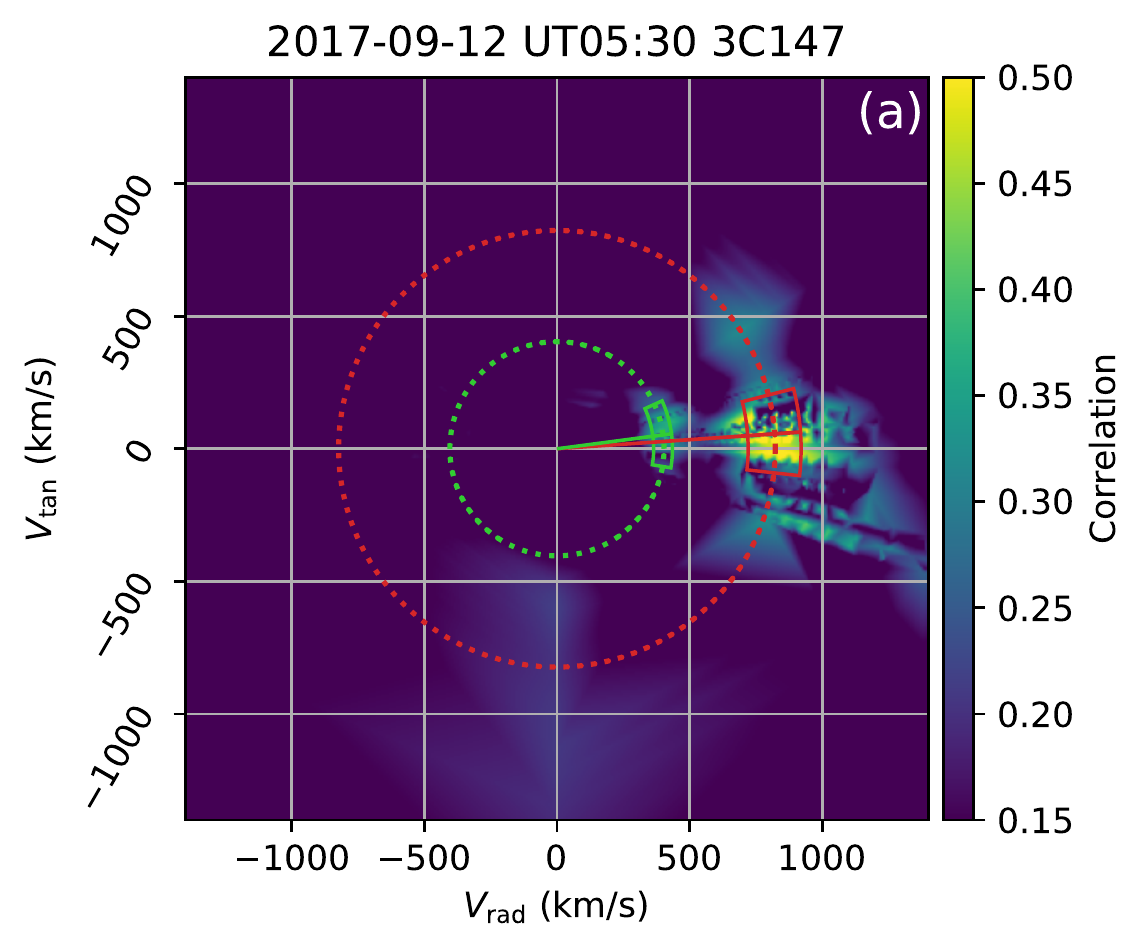}
    \includegraphics[width=\linewidth]{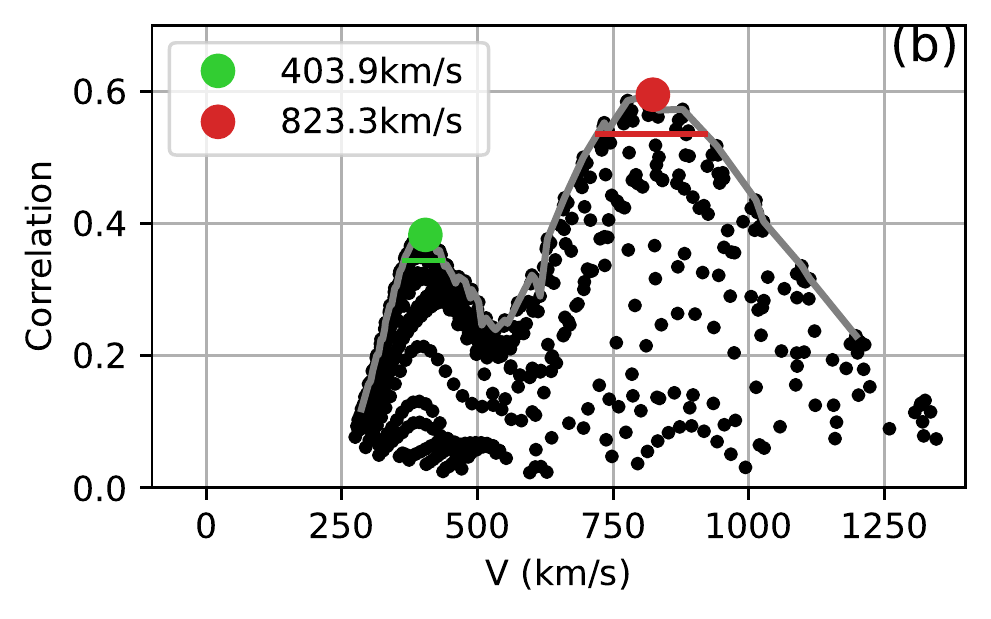}
    \includegraphics[width=\linewidth]{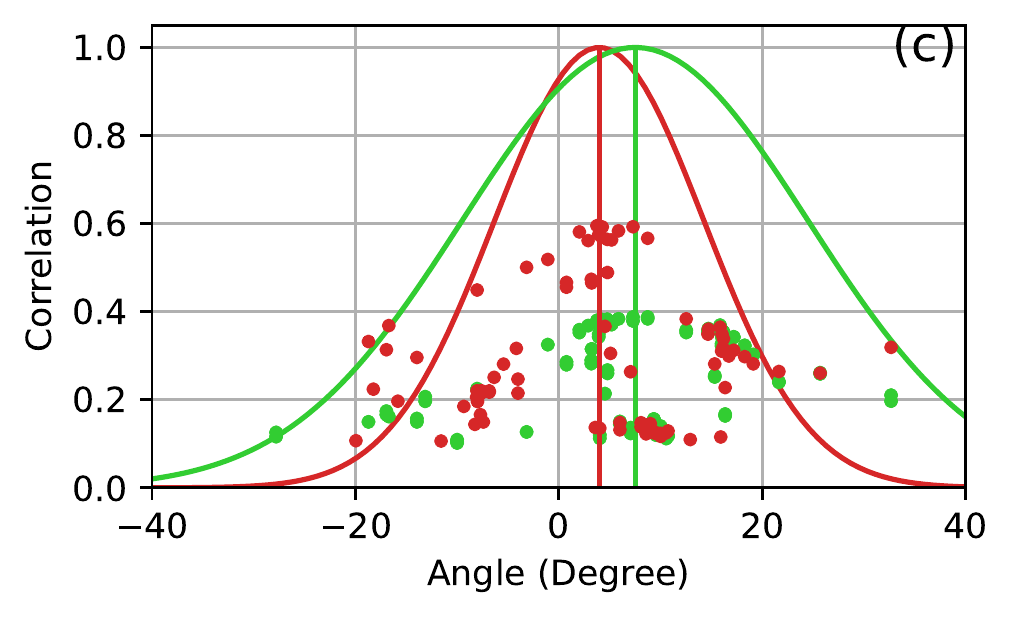}
    \caption{(a) Velocity distribution for the observation of 3C147 at 05:30\,UT on 12 September 2017. (b) Cross-section of the distribution in the radial direction.  (c) Angular distributions along the red and green dotted circles displayed in (a).}
    \label{fig:vdist}
\end{figure}

A further method of visualising velocities is to look at the 2-D velocity distribution.  This technique is still under investigation and not yet robust, so only a first demonstration is presented here to illustrate a further way in which these data can be visualised.  Velocities are calculated for every time-lag in each CCF, for CCF values above 0.1, in both the radial and tangential directions by simple division of the radial and tangential baselines respectively by the time-lag.  These are mapped onto a 2-D velocity space ($V_{rad}$,$V_{tan}$), with each point colour-coded according to the value of the CCF at the time-lag corresponding to that velocity.  This results in a brightness distribution map of velocities which is then displayed as an image as shown in Figure \ref{fig:vdist}(a).  Here it can be seen that there are two blobs, one at a lower velocity representing the background solar wind component, and one covering a range of higher velocities representing the CME component, thus enabling a characterisation of the velocity distribution of multiple components.  Here, the solar wind component shows a well-defined velocity, including direction, whereas the CME component shows a much broader range of values which, speculatively, could reflect the CME propagation across multiple points of the CME front. 

Figure \ref{fig:vdist}(b) presents a cross-section of the distribution in the near radial direction ($\pm5^\circ$), showing the two peaks corresponding to the background solar wind (marked as green) and the CME (marked as red), with velocities of $V_{\rm sw}=$ 404 and $V_{\rm CME}=$ 823\,km\,s$^{-1}$, the widths of which at 90\% peak are 74 and 199\,km\,s$^{-1}$, respectively. These 90\% bounds are shown in Figure \ref{fig:vdist}(a) as the radial width of the green and red boxes respectively.

Figure \ref{fig:vdist}(c) shows the angular distribution of each component, plotting cross-correlation values collected in the velocity range of $\pm 2.5\%$ of $V_{\rm sw}$ and $V_{\rm CME}$, shown as the green and red dotted circles given in Figure \ref{fig:vdist}(a), along with simple single-amplitude Gaussian fits to the data.  It can be seen that these distributions are off-set from the radial direction by 7.5$^\circ$ and 4.0$^\circ$ for the background solar wind and CME respectively. The standard deviation of the fitted Gaussian distribution is also shown in Figure \ref{fig:vdist}(a) presented as the angular width of the green and red boxes.

The values for both velocity and direction are consistent using this technique with those found using the fitting technique above.

\subsection{g-Level}
\label{subsec:glevel}

The amount of scintillation is quantified by a scintillation index, $m$, defined as either the RMS of the scintillation, normalised by the mean intensity,

\begin{equation}
    m = \sqrt{\frac{\langle \Delta I(t)^2 \rangle}{\langle I \rangle^2}}
    \label{eqn:si}
\end{equation}

\noindent or as a simple integration of the filtered and noise-subtracted power spectrum \citep[see, e.g., a summary given in][]{Manoharan:2010},

\begin{equation}
    m^2 = \frac{1}{\langle I \rangle^2} \int_0^{f_c} P(f)df
    \label{eqn:m}
\end{equation}

\noindent which relates it to the spectrum of density fluctuations (Equation \ref{eqn:pspec}).  Here, $f_c$ is the cut-off frequency, i.e., the low-pass filter value used in the processing.  

As a phased-array instrument in which source observation and tracking is achieved through beam-forming the signals from multiple, static, antennas, a further correction to $m$ is required for observations using LOFAR.  The size of the beam formed in the direction of the radio source increases as the source elevation decreases, resulting in a corresponding loss in sensitivity.  Ideally a full beam model incorporating the antennas used in each station beam and their measured beam patterns should be used to fully correct for this effect.  However, using a fit performed to sensitivity versus elevation for pulsar observations \citep{Noutsosetal:2015}, a simple division of $m$ by $sin^{1.39}(elevation)$ acts as a respectable first-order approximation and this is what is applied to the LOFAR data used here.

$m$ increases under weak scattering conditions with decreasing distance towards the Sun, before reaching a peak and decreasing rapidly under strong scattering conditions.  Therefore, many measurements taken over a range of solar elongations for a given radio source are normally used to establish a basic $m$-elongation relation which is then used to normalise for this effect.  The normalisation also removes the effect of source structure, and the result of it is an index known as the $g$-level.

\begin{figure}
    \centering
    \includegraphics[width=\linewidth]{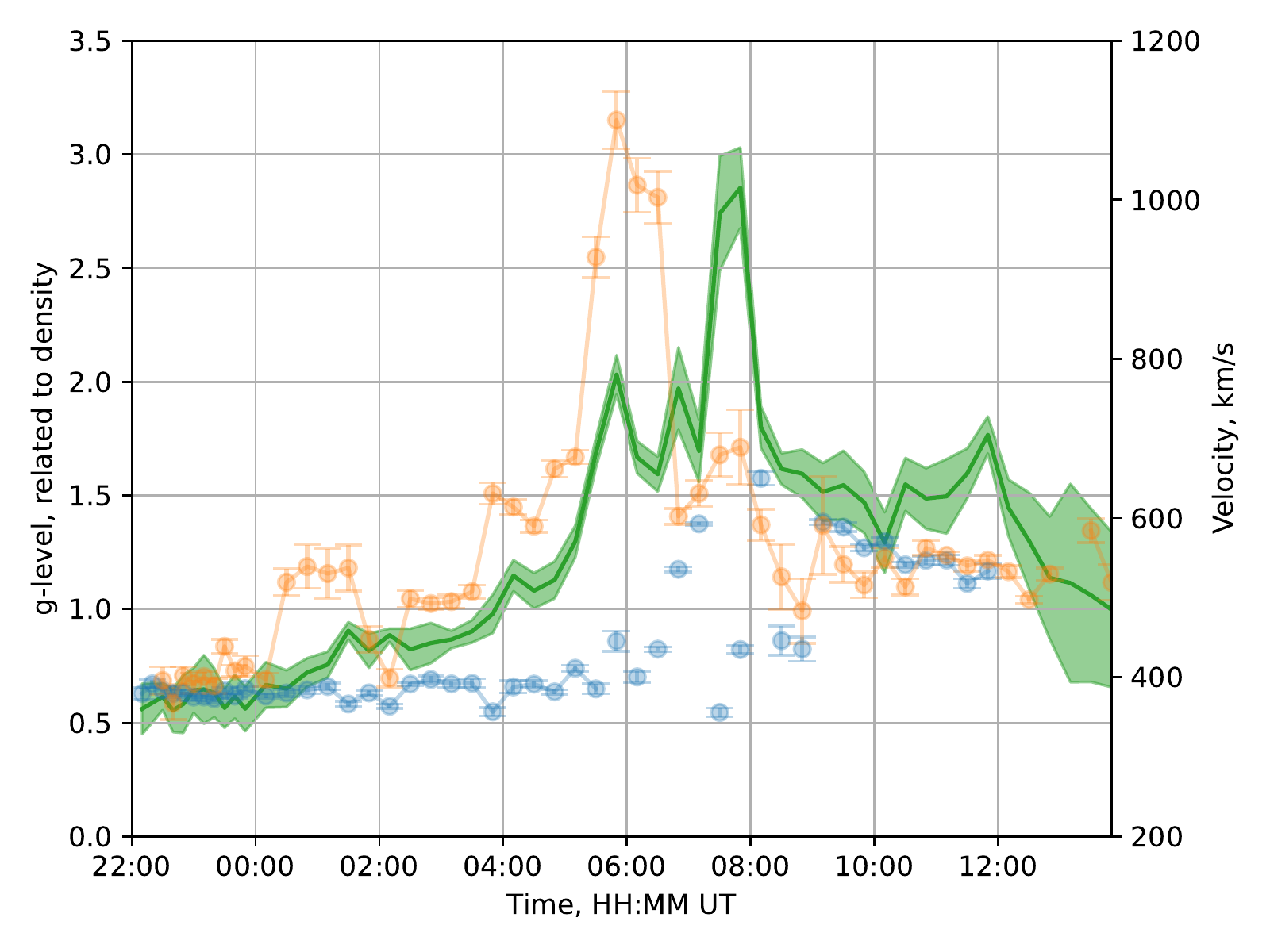}
    \caption{Estimated g-levels for observations of 3C147 taken on 12 September 2017 (green). The velocities from Figure \ref{fig:velocities} are plotted in the background for comparison.}
    \label{fig:glevel}
\end{figure}

The regular analysis detailed in Section \ref{sec:observations} results in the  calculation of power spectra, filtered as described, and a noise floor subtracted.  Scintillation indices were therefore taken as the simple integration of the power spectra, using the square root of Eqn. \ref{eqn:m}.  Since LOFAR has yet to perform the measurements necessary for a proper normalisation to calculate $g$-level, the scintillation indices were divided by the mean of those taken between 22:00\,UT and 23:00\,UT on 11 September 2017 and the results taken to be an estimated $g$-level.  A further correction was applied to match g-levels at the start of the period used (22:00\,UT on 11 September 2017) with the value measured for 3C147 by ISEE at 21:00\,UT on 11 September 2017, as used by \citet{Iwaietal:2022}.  The results are plotted in Figure \ref{fig:glevel}, where the green line is the median of g-levels calculated for intensities received by each station, and the range given as the median absolute deviation.  Plotted in the background are the velocities displayed in Figure \ref{fig:velocities} for comparison.

It can be seen that the estimated $g$-level undergoes an initial rise corresponding with the main peak in velocity seen in Figure \ref{fig:velocities}.  A further substantial rise is seen approximately two hours later, corresponding with a slight bump in the velocity. Thereafter, the g-level remains elevated but trending generally downwards.

\subsection{Spatial Correlation}
\label{subsec:spatialcorrelation}

The number of stations available to LOFAR means that it can be used to obtain direct visual interpretations of the component parts expressed in equations \ref{eqn:pspec} to \ref{eqn:phine}, and thereby determine from imaging fundamental details about the density fluctuations themselves, by effectively using it as an intensity interferometer.  The main integration part of Equation \ref{eqn:pspec} describes the two-dimensional \textit{spatial} spectrum of the density fluctuations, convolved with the effects of radio source structure (Equation \ref{eqn:fsource}) and the Fresnel propagation filter (Equation \ref{eqn:Fresnel}).  This is converted into a temporal spectrum by the fact that the spatial spectrum is moving with the solar wind velocity.  The inverse FFT of the spatial spectrum is the spatial correlation function, and this is directly sampled by all the CCFs of an observation of IPS using a variety of baseline lengths and orientations.

\begin{figure}
    \centering
    \includegraphics[width=\linewidth]{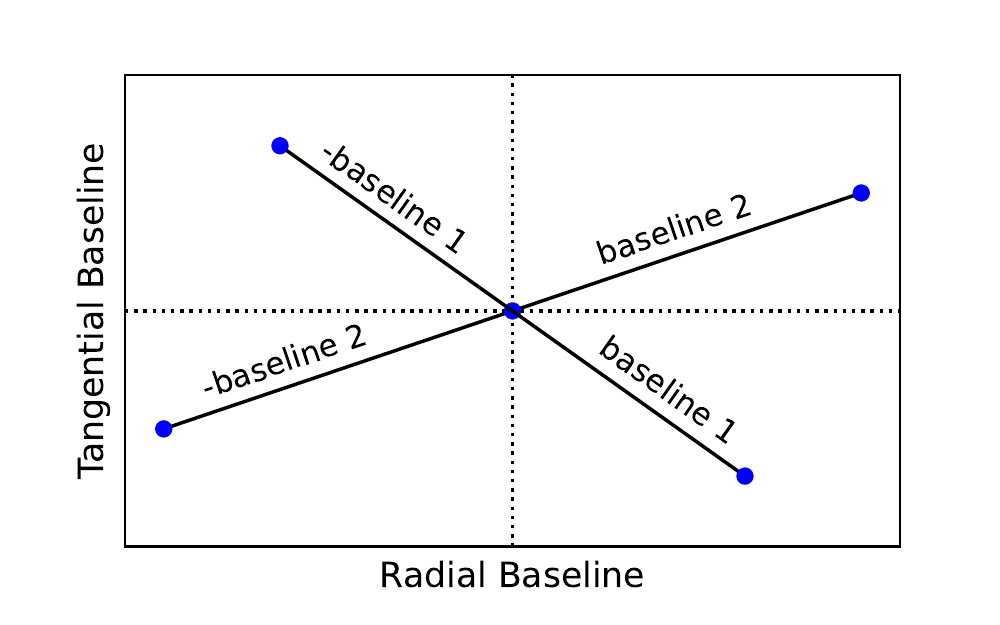}
    \caption{Diagram illustrating how spatial correlation images are built up.}
    \label{fig:spatialbaselines}
\end{figure}

The application of this to the data can be illustrated using the values of the CCFs at zero time-lag (0\,s).  The zero-lag value of each CCF is placed on a spatial (radial,tangential) grid at both the point equating to the radial and tangential baseline components, and at their negative equivalent, as illustrated in Figure \ref{fig:spatialbaselines}.

\begin{figure}
    \centering
    \includegraphics[width=\linewidth]{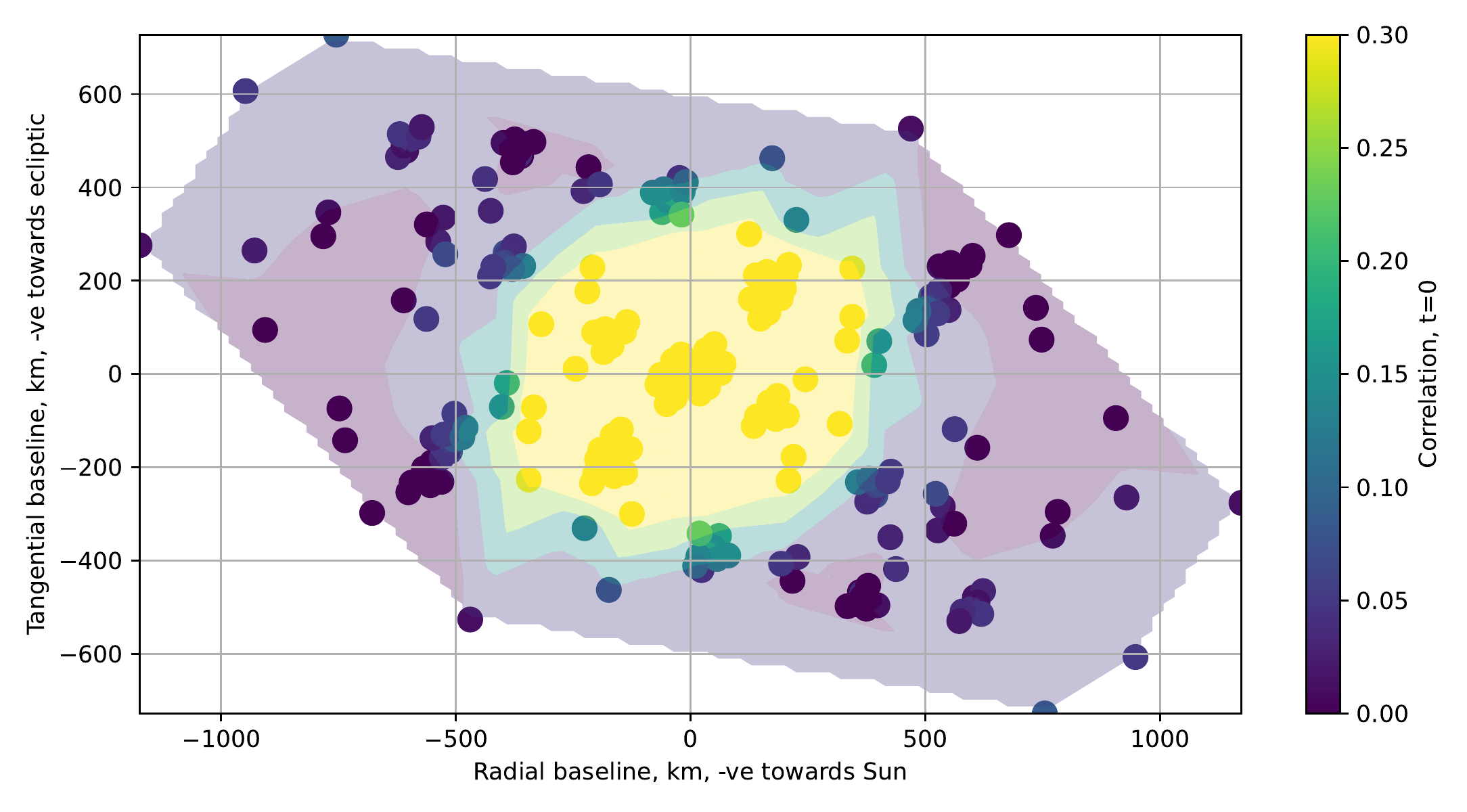}
    \caption{Scatter plot on a (radial,tangential) baseline grid of the zero time-lag cross-correlation values for the observation of 3C147 at 01:50\,UT on 12 September 2017.  Values are placed at both positive and negative equivalents of each baseline, as described in the text.  Superposed are filled contours fitted to these values.}
    \label{fig:zerolag}
\end{figure}

The result is a scatter plot of data points, as shown in Figure \ref{fig:zerolag}, which is then contoured and the contours filled to bring out of the overall shape of the structure.  In all subsequent plots this is displayed only as an image of the contour data, with values clipped to emphasise the main structure.  Since the Fresnel filter (Equation \ref{eqn:Fresnel}) simply places a limit on the scale size of the density fluctuations, the shape and orientation of the spatial structure displayed here is basically a reflection of the shape and orientation of the density structure as described in Equations \ref{eqn:phine} and \ref{eqn:q} (which would indicate the axial ratio of the density fluctuations and magnetic-field alignment under the assumption that the fluctuations are indeed aligned with this), convolved with the shape and orientation of the radio source itself (Equation \ref{eqn:fsource}).

\begin{figure}
    \centering
    \includegraphics[width=\linewidth]{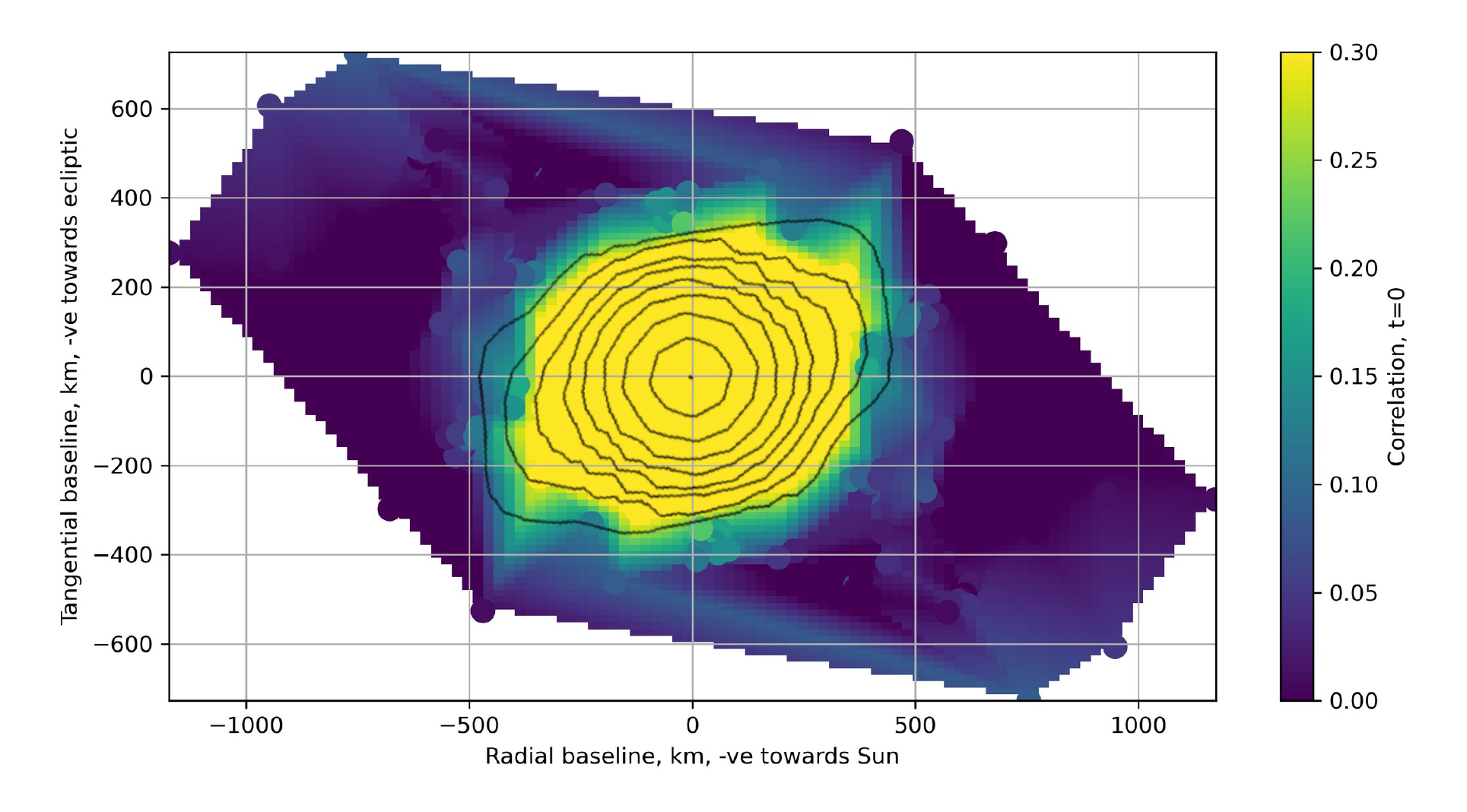}
    \caption{Contours showing the structure of 3C147, as observed using MERLIN at 151\,MHz and taken from Figure 1(d) in \citet{Akujoretal:1990} scaled maintaining the correct aspect ratio, and oriented to correct for the angle between Right Ascension and the solar radial direction at this time, to fit an image of the LOFAR spatial correlation structure from Figure \ref{fig:zerolag} underneath.}
    \label{fig:source}
\end{figure}

Radio source 3C147 is, unfortunately, not an ideal point source and so its structure must be accounted for if any analysis is to obtain accurate information on the small-scale density structure of the out-flowing plasma.  This radio source has a slightly elongated structure which bears some similarity to the elongated structure seen in Figure \ref{fig:zerolag}.  Figure \ref{fig:source} shows an image of the contour fit shown in Figure \ref{fig:zerolag}, but with contours taken from Figure 1(d) in \citet{Akujoretal:1990} showing the structure of 3C147 at 151\,MHz (in the centre of the LOFAR frequency band used for the IPS observations here) superposed.  The radio source contours have been scaled to match the LOFAR image while maintaining a correct aspect ratio, but the axes on the plot do not reflect any physical parameter describing the size of the radio source itself.  The radio source plot was also rotated to correct for the angular difference between Right Ascension and the solar radial direction at this time.  From Figure \ref{fig:source} it can be inferred that the zero-lag spatial structure shown appears primarily to be a reflection of the radio source itself in this instance.

\begin{figure}
    \centering
    \includegraphics[width=\linewidth]{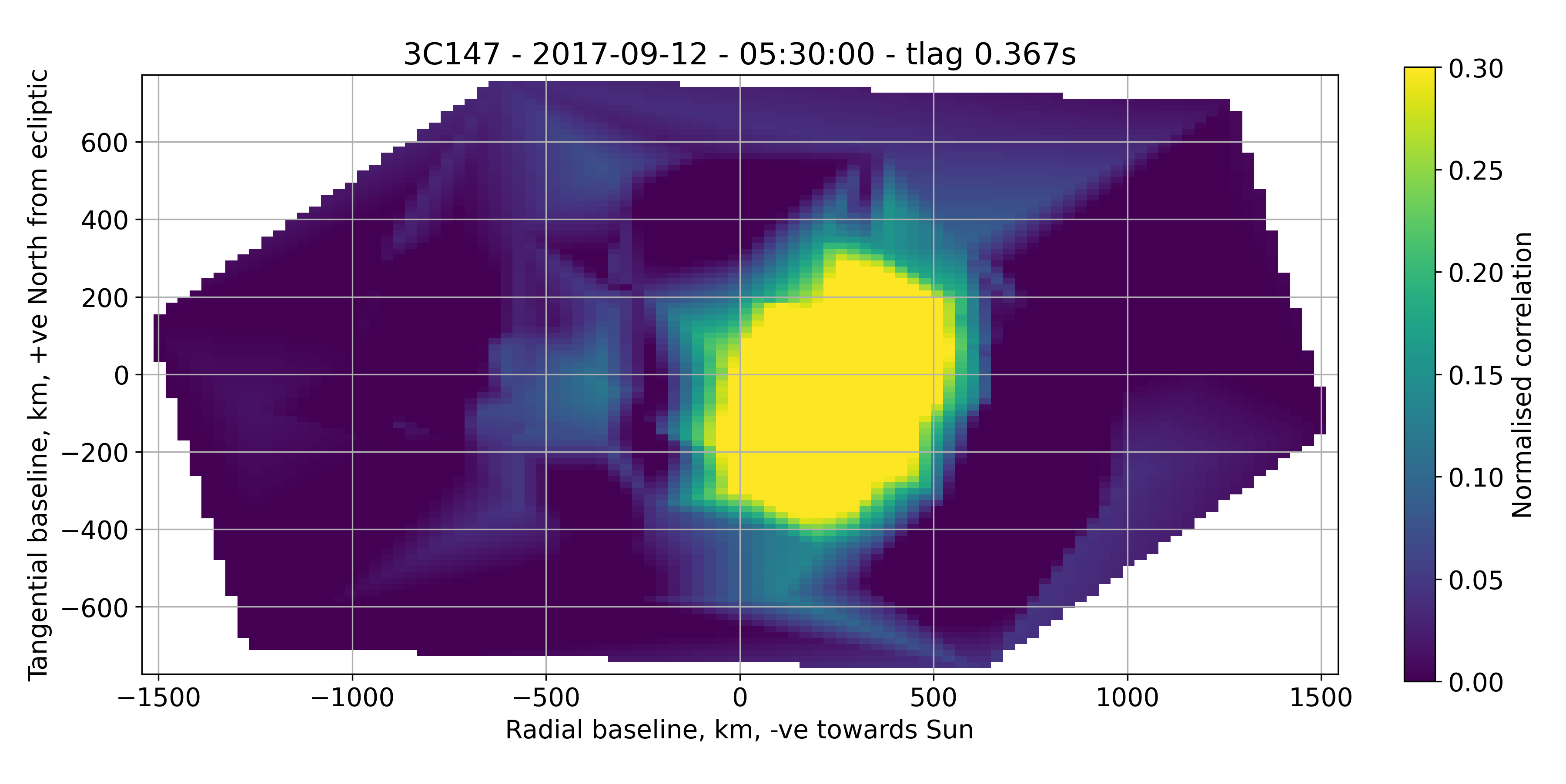}
    \includegraphics[width=\linewidth]{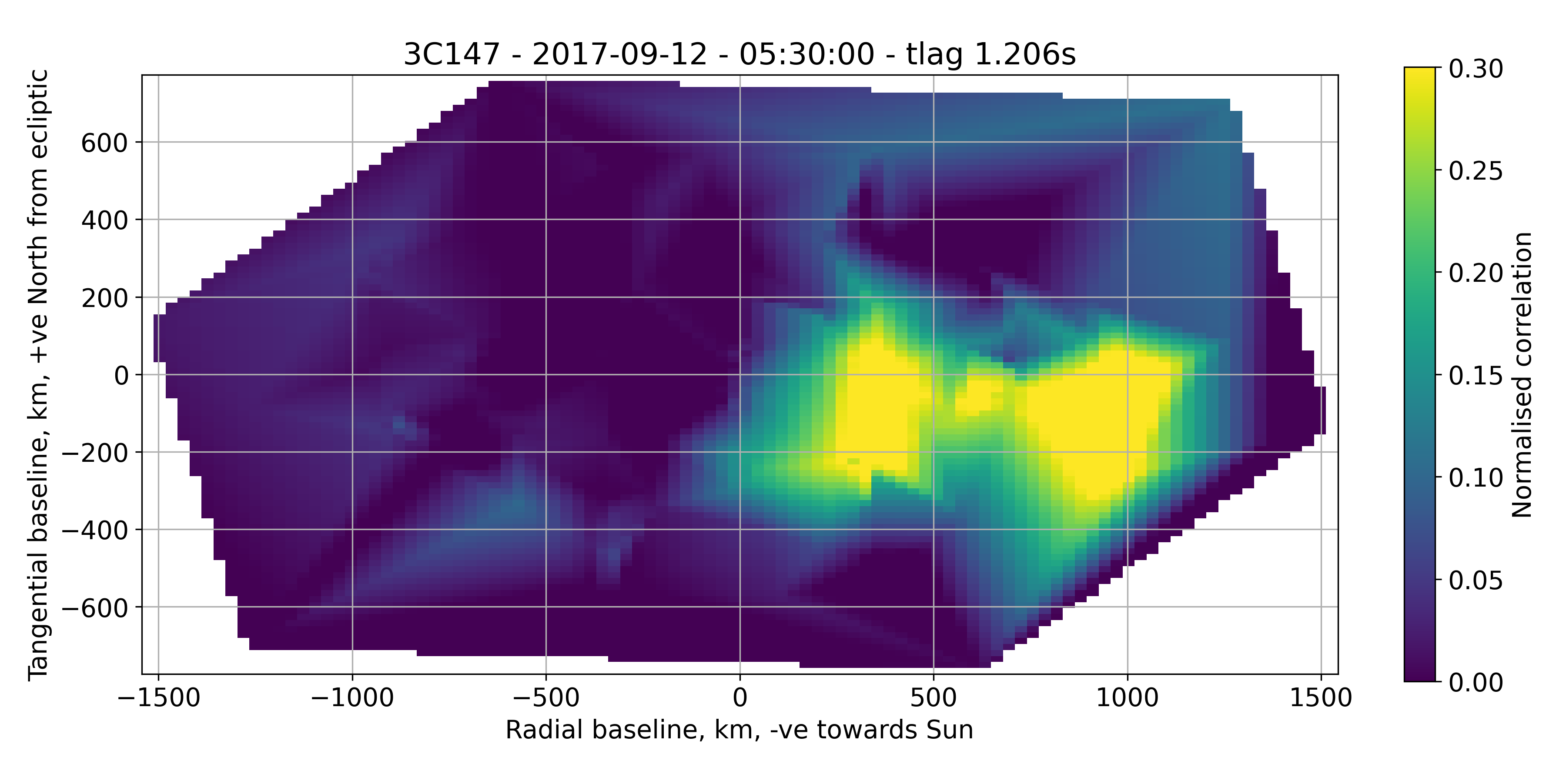}
    \includegraphics[width=\linewidth]{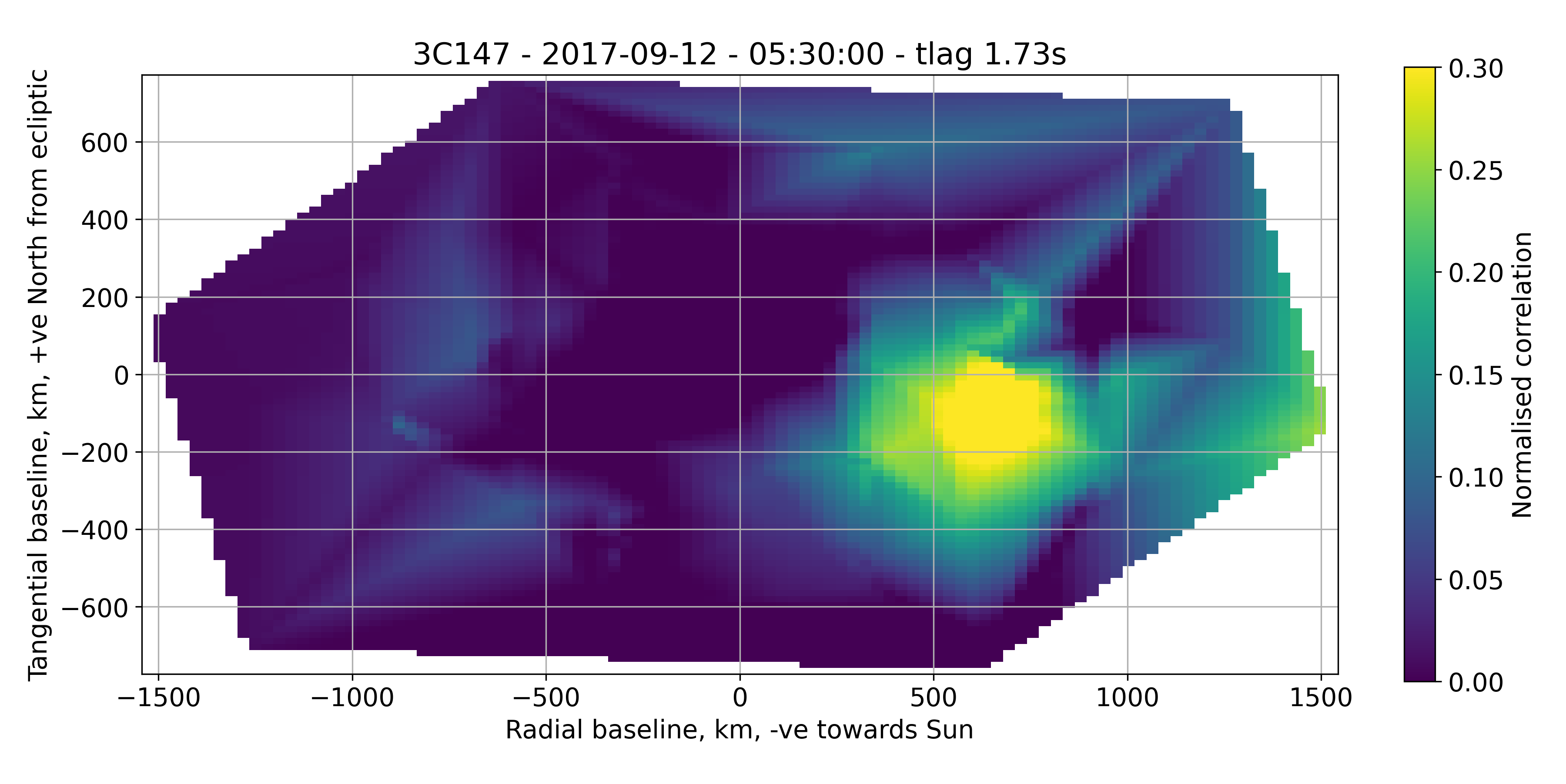}
    \includegraphics[width=\linewidth]{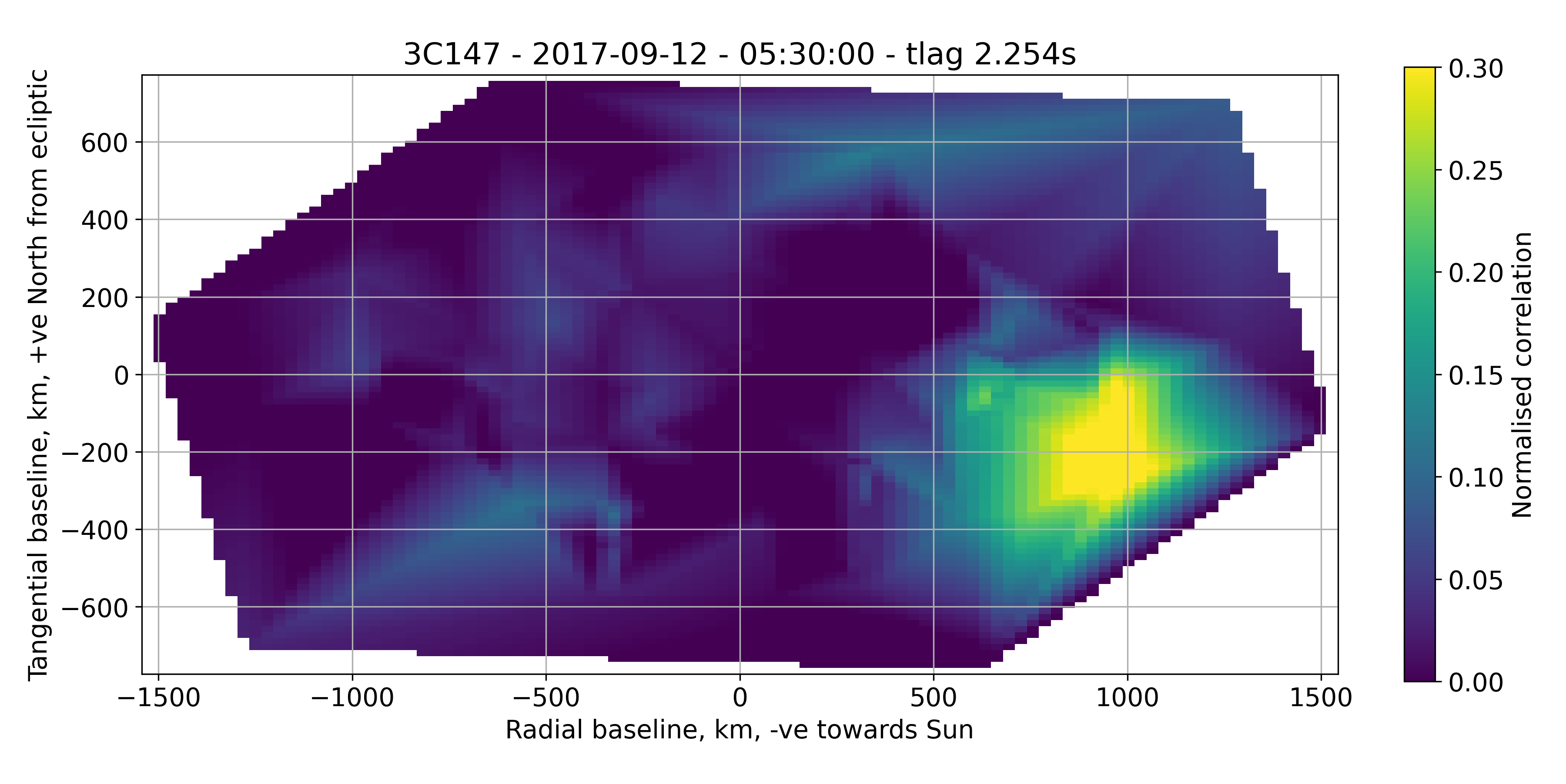}
    \caption{Spatial correlation images for different time-lags of the CCFs calculated for the observation of 3C147 at 05:30\,UT on 12 September 2017; (a) 0.367\,s, (b) 1.206\,s, (c) 1.730\,s, and (d) 2.254\,s.  Each image is of the contours fitted to the correlation values, as in Figure \ref{fig:zerolag}, clipped to a maximum value of 0.3 to more clearly indicate the shape.  The x-axis represents the radial direction, and the y-axis the tangential direction.}
    \label{fig:spatialtimelags}
\end{figure}

The remainder of the CCF contains a great deal of additional information when viewed in spatial plots such as these.  For any time-lag the corresponding cross-correlation values can be plotted on the same spatial grid as the zero-lag plots, associating the baselines taken in a positive radial direction with the values at a positive time-lag, and plotting the values at the equivalent negative time-lag in the equivalent negative baseline direction.  By cycling through all the time-lags, a movie of the spatial cross-correlation can be made, which is provided in online material.  Figure \ref{fig:spatialtimelags} shows four segments from this movie, for four different time-lags.

Figure \ref{fig:spatialtimelags} shows the spatial pattern moving and separating into two distinct components, the faster component representing the small-scale density structure of the CME, and the slower component that of the background solar wind.  Although not obvious in this particular data set, the structure of the radio source remains at zero time-lag as it is a constant across the data received by all stations.  The initial structure (Figure \ref{fig:spatialtimelags}(a)) is dominated by the CME: it appears elongated in a direction almost perpendicular to the radial, suggesting that this direction may be the alignment of the magnetic field at this point.  The CME component then properly separates out at longer time-lags (Figure \ref{fig:spatialtimelags}(b) - the small additional component between the two larger ones is static and therefore likely to be an artifact of the data), leaving the background solar wind component moving in a slightly off-radial direction.

\begin{figure}
    \centering
    \includegraphics[trim=150 0 90 0,clip,width=0.28\linewidth]{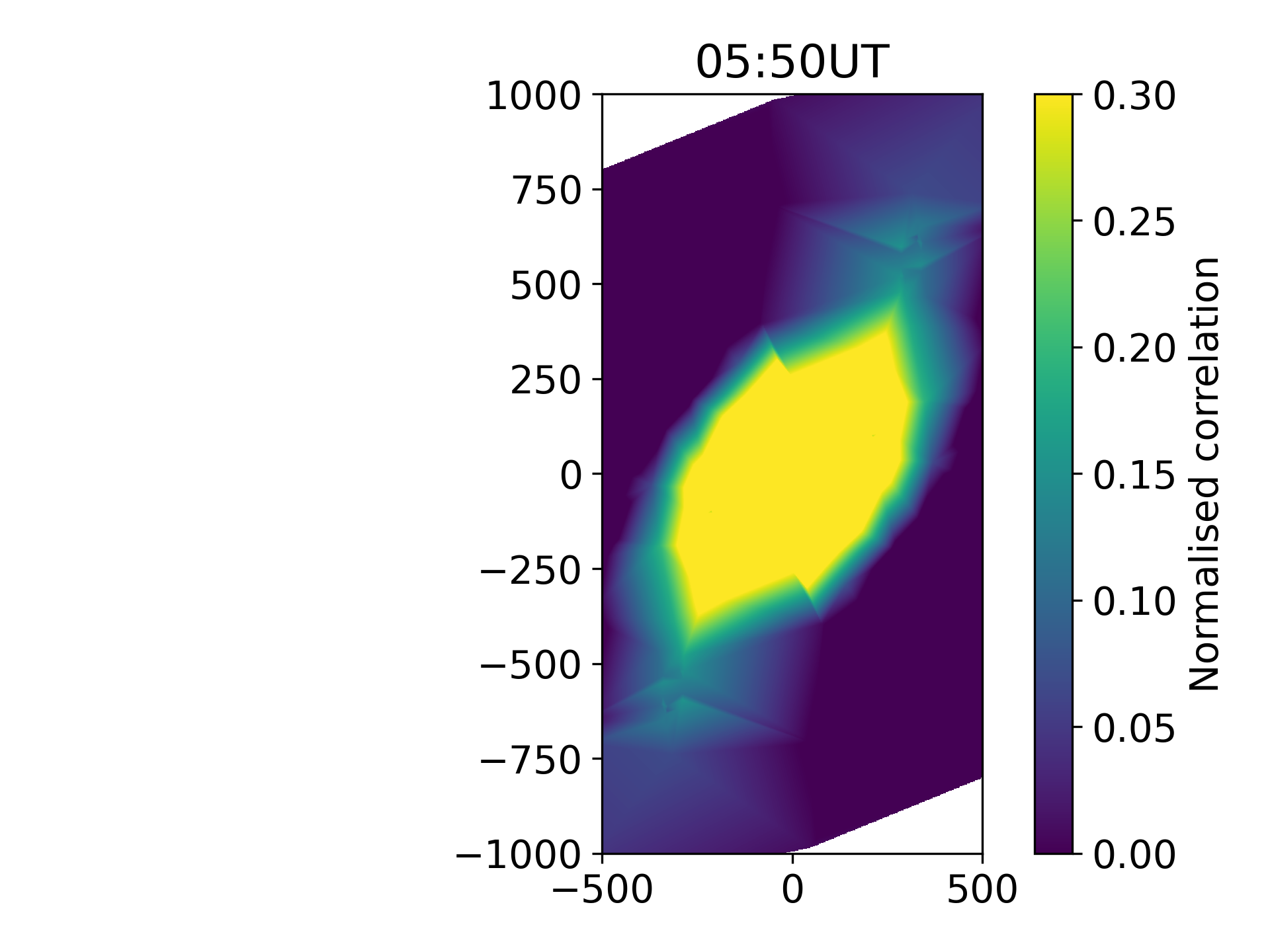}\includegraphics[trim=150 0 90 0,clip,width=0.28\linewidth]{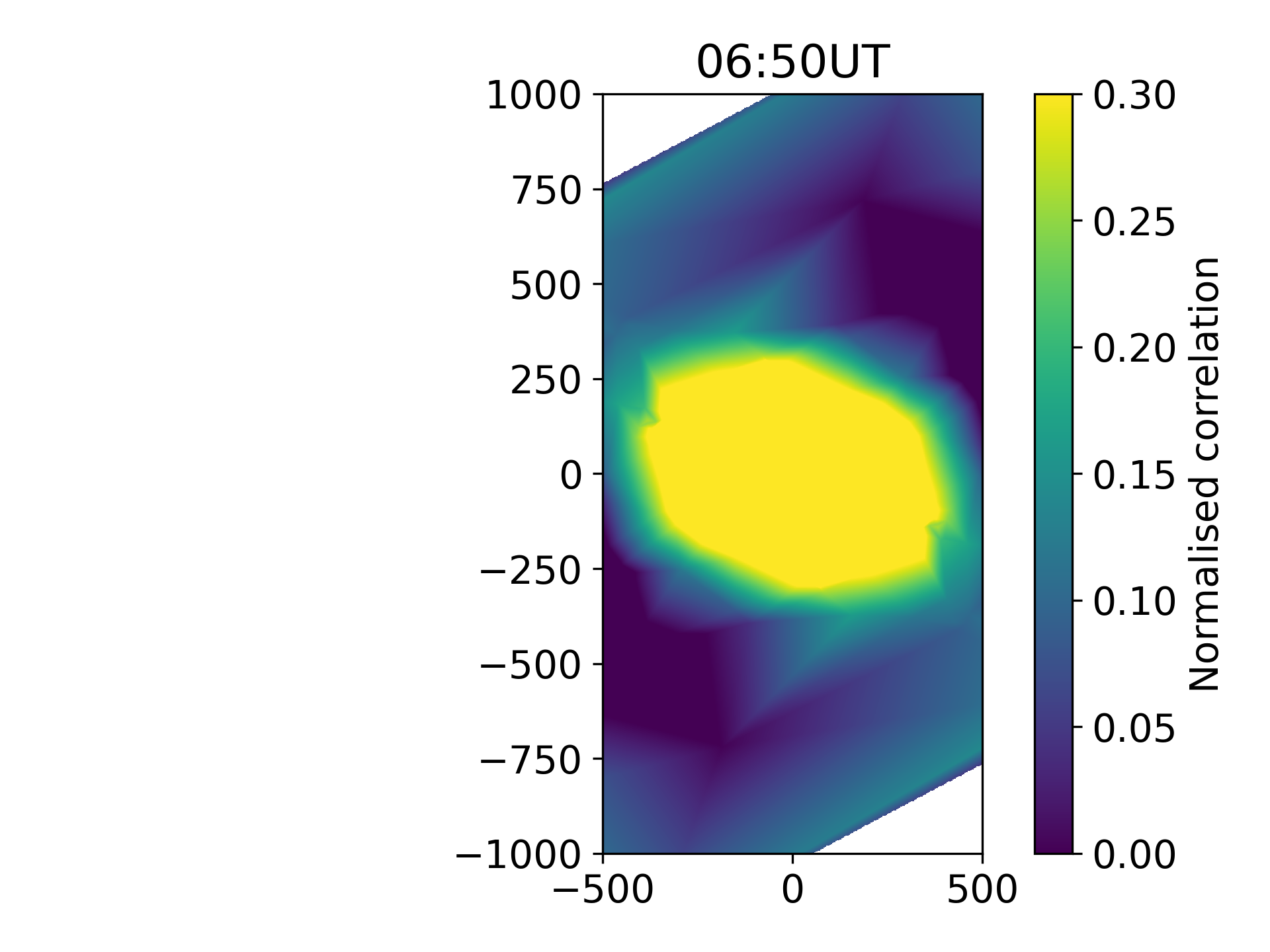}\includegraphics[trim=150 0 0 0,clip,width=0.4\linewidth]{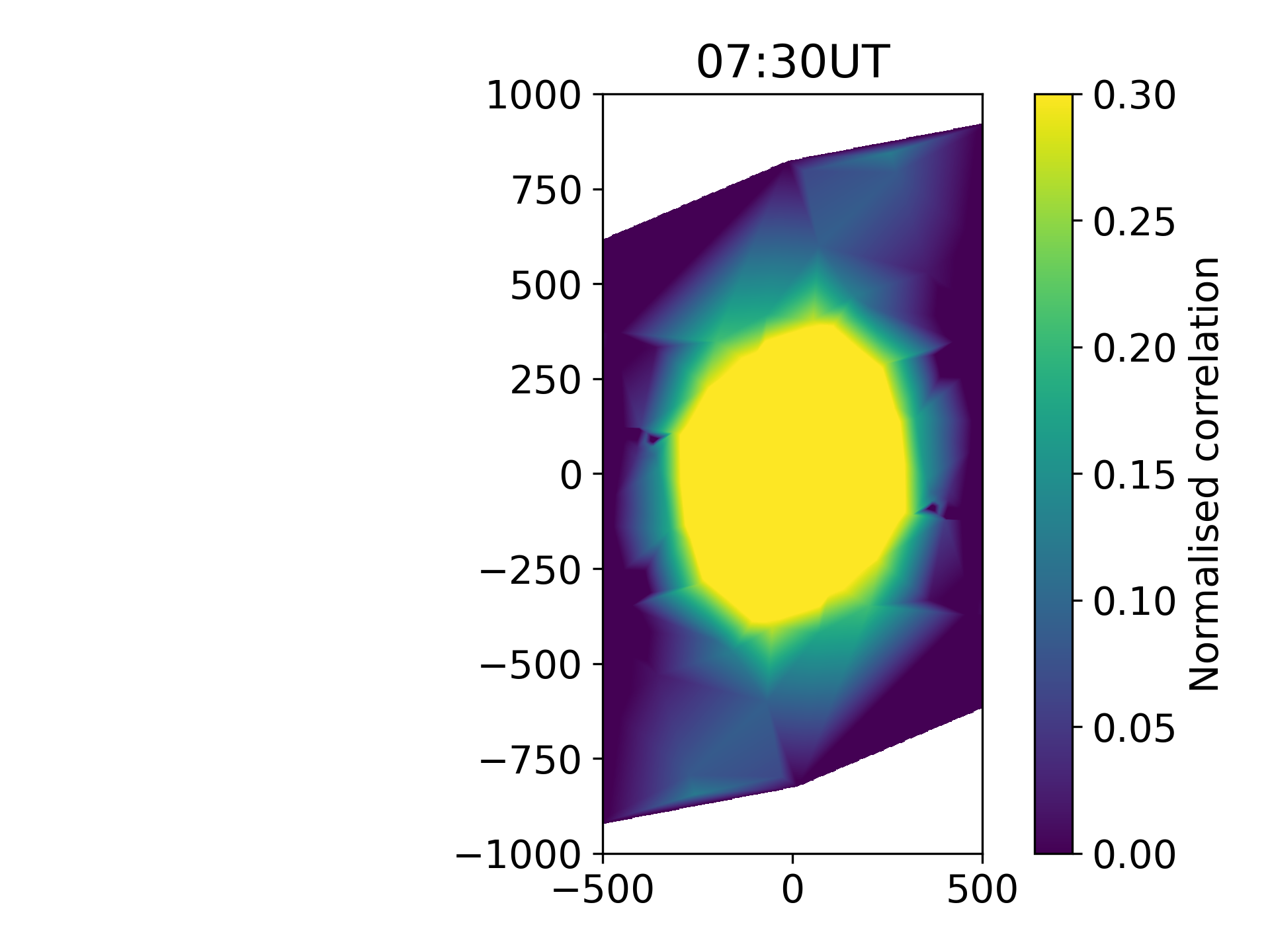}
    \caption{Spatial correlations at zero time-lag for three different times during the passage of the merged CME event.}
    \label{fig:spatialdirections}
\end{figure}

\begin{figure}
    \centering
    \includegraphics[width=\linewidth]{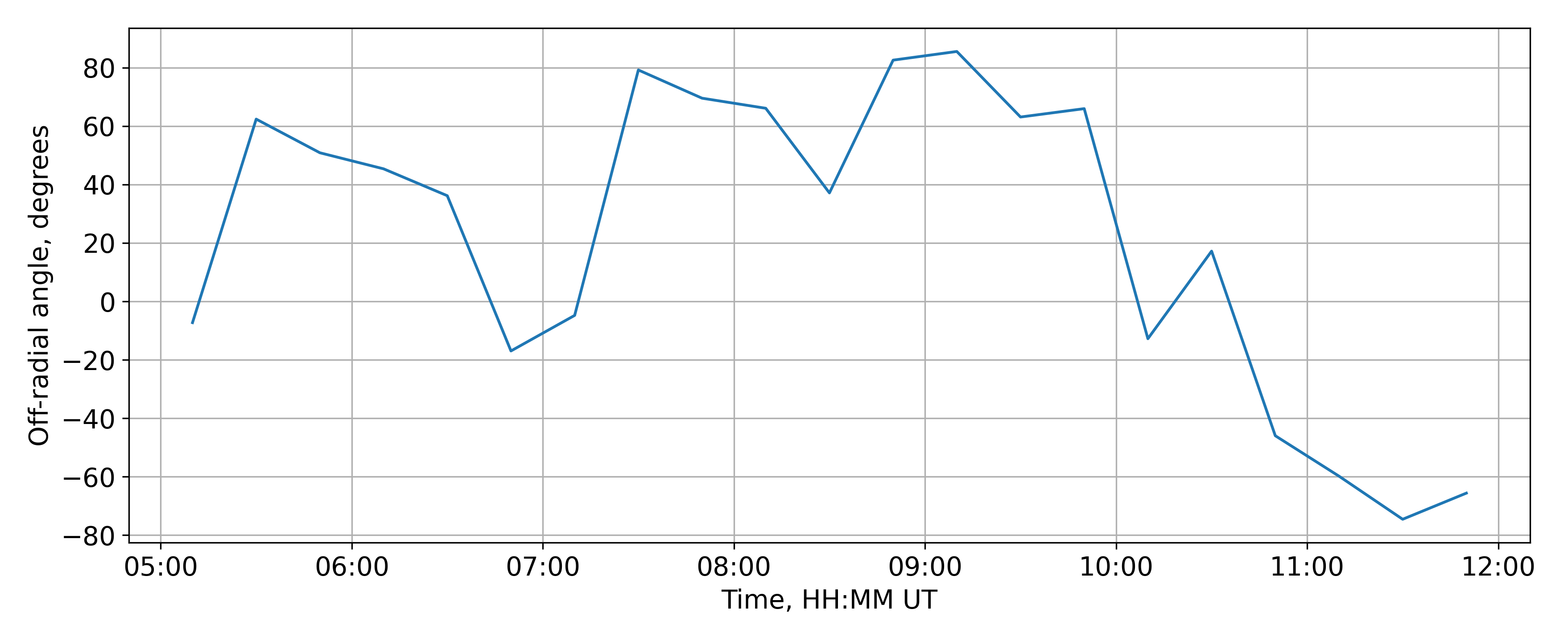}
    \caption{Orientations of the elongated structure seen in the spatial correlation images during the course of the passage of the merged CMEs.}
    \label{fig:cmeangles}
\end{figure}

The orientation of the small-scale density structure changes during the course of the observations, as demonstrated in Figure \ref{fig:spatialdirections}.  The angle of this orientation to the radial direction can be calculated by fitting a 2-D Gaussian to the spatial structure and measuring the fitted orientation of the long axis.  The resulting values, corrected where necessary such that only angles in the positive radial quadrants are given (i.e. orientations in the positive tangential baseline direction are positive, and vice versa), are plotted in Figure \ref{fig:cmeangles}.

Since the orientation of the radio source hardly changes during the course of these observations, these orientations represent estimates of the orientation of the anisotropy of the density fluctuations (equation \ref{eqn:q}), itself reflecting the orientation of the magnetic field.  The results presented in Figure \ref{fig:cmeangles} are for the times when the observations were dominated by the CME passage, and so most likely reflect the orientation of the CME magnetic field, projected onto a plane perpendicular to the line of sight.

\section{Discussion}

\begin{figure}
    \centering
    \begin{subfigure}{\linewidth}
        \centering
         \includegraphics[trim=0 10 0 20,clip,width=0.8\linewidth]{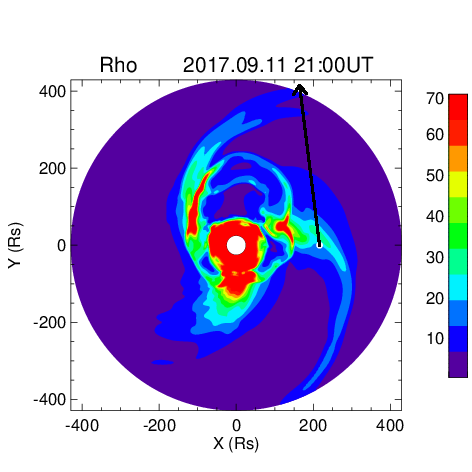}
        \caption{11 September 2017, 21:00\,UT.}
        \label{subfig:susanooisee0}
    \end{subfigure}
    \begin{subfigure}{\linewidth}
        \centering
         \includegraphics[trim=0 10 0 20,clip,width=0.8\linewidth]{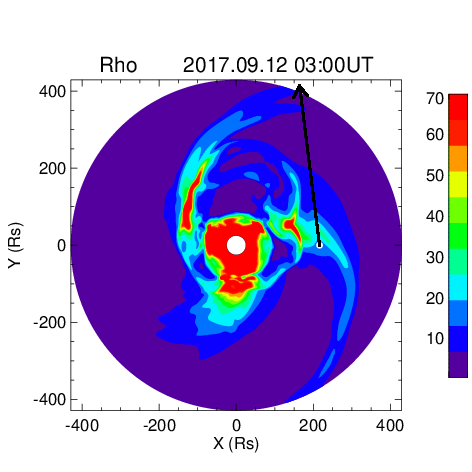}
        \caption{12 September 2017, 03:00\,UT.}
        \label{subfig:susanooisee1}
    \end{subfigure}
    \begin{subfigure}{\linewidth}
        \centering
         \includegraphics[trim=0 10 0 20,clip,width=0.8\linewidth]{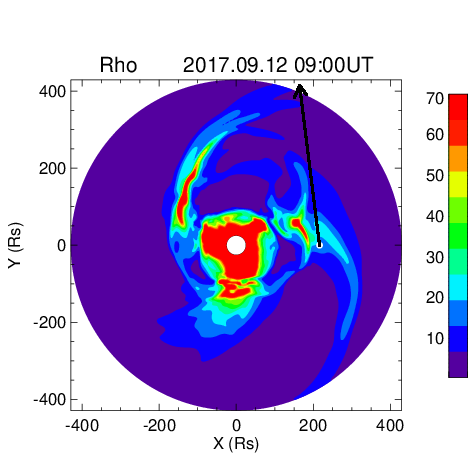}
        \caption{12 September 2017, 09:00\,UT.}
        \label{subfig:susanoolofar}
    \end{subfigure}
    \caption{Ecliptic cuts through the simulated density distribution using the SUSANOO-CME IPS-based MHD model.  The black arrow represents the direction of the line of sight to 3C147 projected onto the ecliptic plane.  Axis coordinates are given in solar radii, Rs.}
    \label{fig:susanoo}
\end{figure}

The event observed by LOFAR is the result of a merger between two slow CMEs which launched on 9 September 2017, and the ultra-fast CME launched on 10 September 2017 \citep[e.g.][]{Guoetal:2018,Leeetal:2018}.  This merger has also been modelled using the IPS-based MHD model SUSANOO-CME of \citet{Iwaietal:2019}, with the results compared with the g-level measurements from both ISEE and LOFAR, presented in the companion paper by \citet{Iwaietal:2022}.  Figure \ref{fig:susanoo} shows ecliptic cut snapshots of the SUSANOO-CME simulations for 21:00\,UT on 11 September 2017, and 03:00\,UT and 09:00\,UT on 12 September 2017 (times chosen to match the standard times of tomographic reconstructions, presented later), along with projections of the line of sight to 3C147, to illustrate how the model shows the evolution and approach of CME material to the LOFAR observations.

The LOFAR g-levels (Figure \ref{fig:glevel}) show a rise with time which generally follows, with a lag, the main increase in velocity, with the g-level peaking approximately two hours after the peak in velocity.  These high $g$-levels indicate substantially increased turbulence (and/or increased variation in density) associated with the merged CME, which could be associated with the dense material of the merged CME itself crossing the lines of sight and/or from a shock front ahead of it.  The SUSANOO-CME results show a dense co-rotating structure encompassing the Earth and near-Earth part of the 3C147 line of sight, which is likely the source of the slow background solar wind observed by LOFAR before the CME.  The part of the CME approaching the projected 3C147 line of sight appears as a relatively narrow (in the direction of propagation) ribbon of material with a denser core.  The figure (and the full results of \citet{Iwaietal:2022}) also demonstrates that the CMEs approaching the lines of sight to 3C147 were only partially merged, leading to a complex density distribution in parts.  The steady rise in LOFAR g-level suggests an increasing pile-up of material before the main passage of the CME, although timing between the model and LOFAR data is imperfect.






The velocity results (Section \ref{subsec:velocity}) show the first effects of the CME shortly after 00:00\,UT on 12 September 2017, but fade from the scintillation signal for a short period around 02:00\,UT, before a substantial rise in velocity to a peak at around 06:00\,UT, as shown in Figure \ref{fig:velocities}.  The initial increase in velocity at around 01:00\,UT is puzzling: while some short-baseline CCFs show a clear high velocity here and appear reliable, most show a typical slow solar wind stream velocity with little obvious indication of anything faster.  It may also be that the high velocities around 01:00\,UT are significantly faster than those indicated in the figure: those short-baseline CCFs which do show a high velocity indicate speeds of around 800\,km\,s$^{-1}$, but only with a direction some 30$^{\circ}$ off-radial in an equatorial direction, which seems unlikely.  It is possible that this first peak in velocity corresponds to material which is either of similar (or not much higher) density to the background solar wind, or crossing the tails of the lines of sight with a great enough density to be visible, barely, but there is little indication of this in the SUSANOO-CME model results. 

\begin{figure*}
    \centering
    \includegraphics[width=0.4\linewidth]{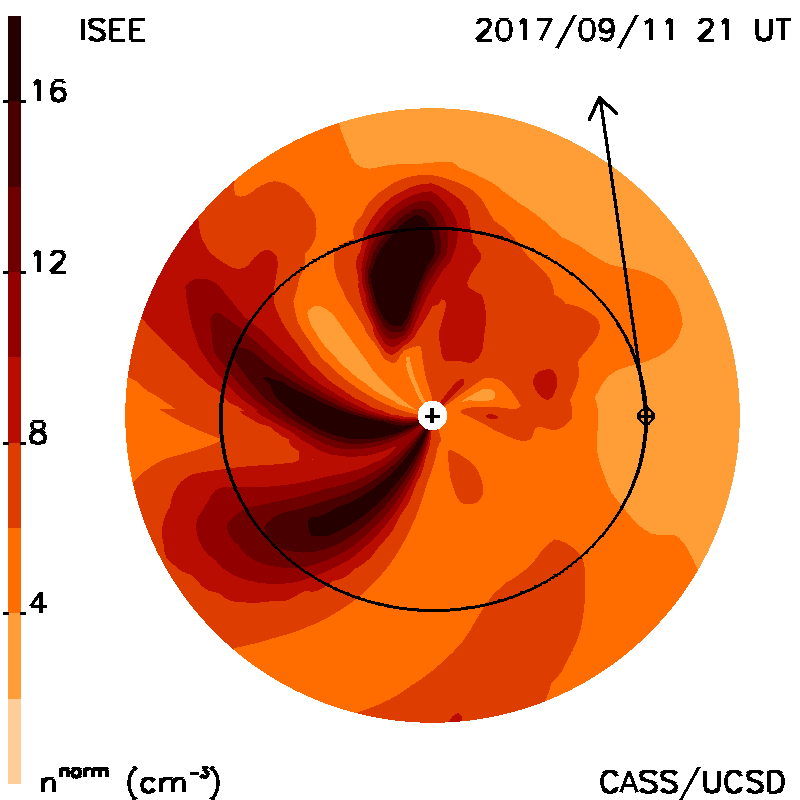}\includegraphics[width=0.4\linewidth]{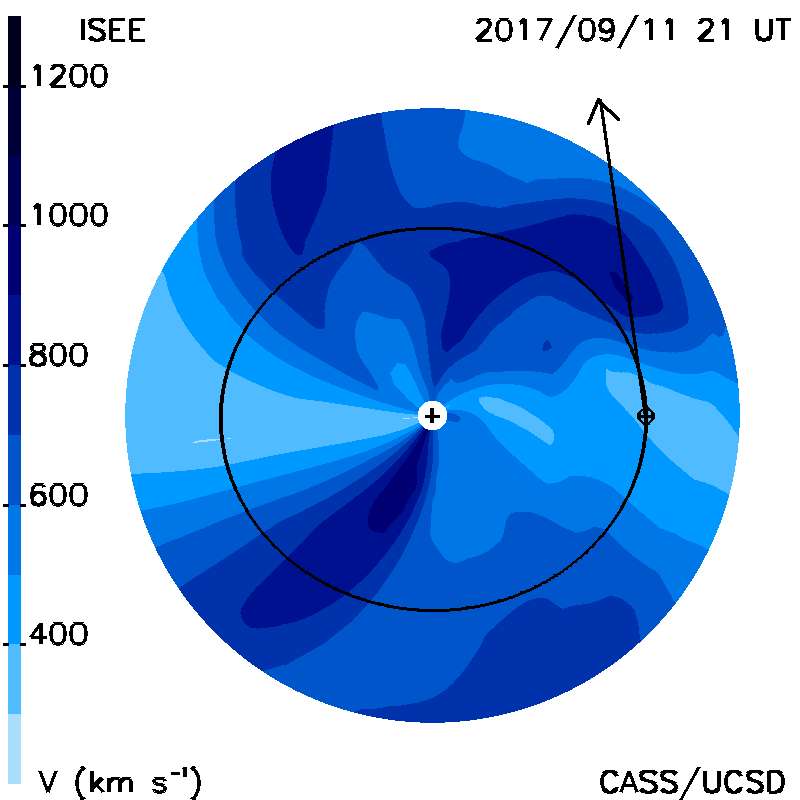}
    \includegraphics[width=0.4\linewidth]{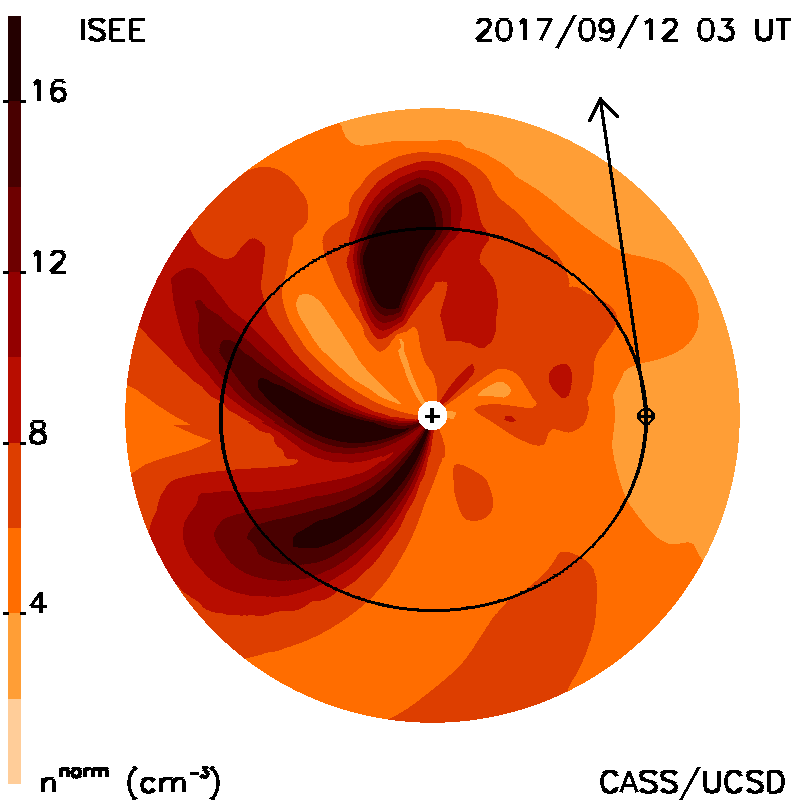}\includegraphics[width=0.4\linewidth]{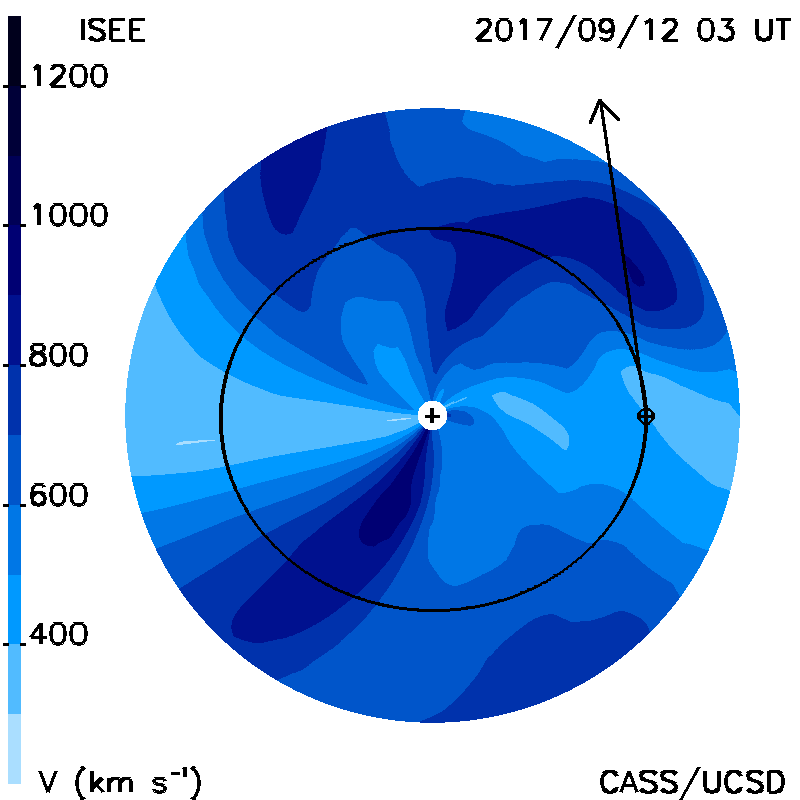}
    \includegraphics[width=0.4\linewidth]{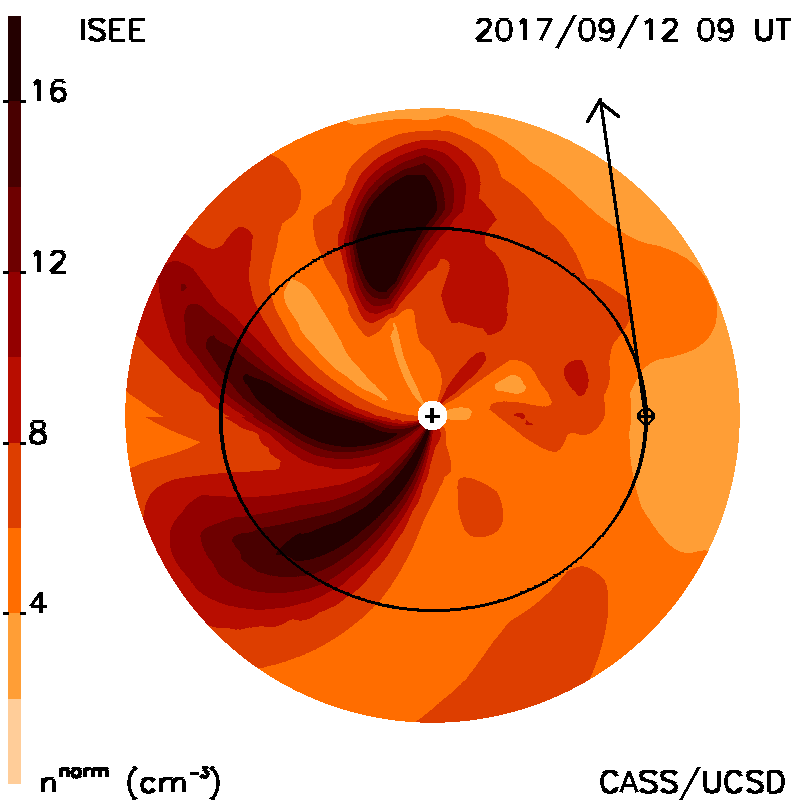}\includegraphics[width=0.4\linewidth]{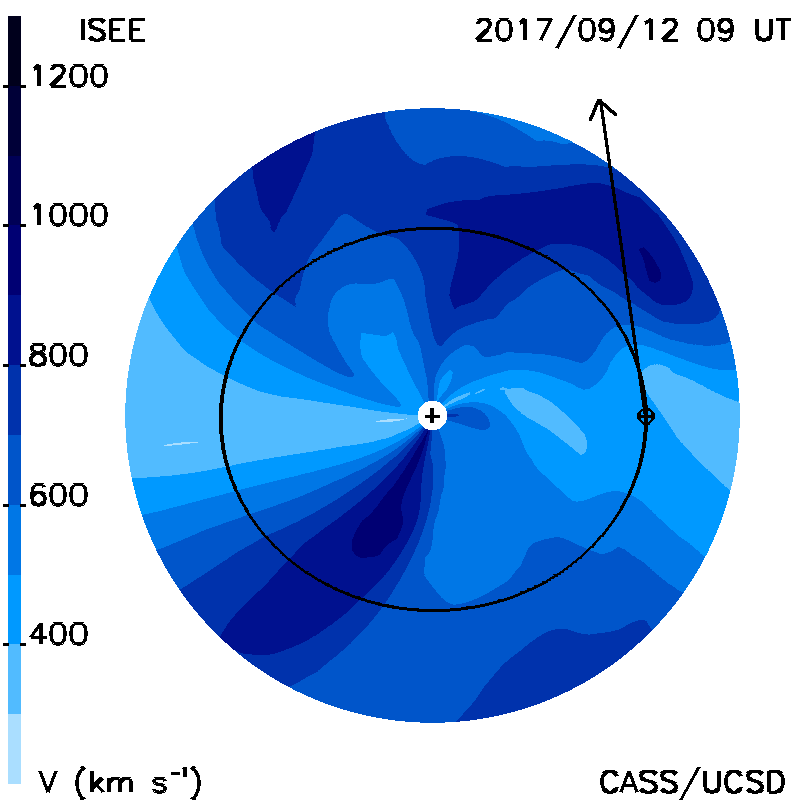}
    \caption{Cuts through ISEE IPS tomographic reconstructions of density (left) and velocity (right) for 21:00\,UT on 11 September 2017 (top), 03:00\,UT on 12 September 2017 (middle), and 09:00\,UT on 12 September 2017 (bottom).  Each cut is in a plane running through the Sun-Earth line and tilted anti-clockwise if viewing the Sun from the Earth to align with the line of sight to 3C147 (the direction shown by the black arrows).  A circular representation of the Earth's orbit is projected onto this plane (black oval), with the Earth shown on the right-hand side.}
    \label{fig:tomo}
\end{figure*}

Tomographic reconstructions of density and velocity using ISEE IPS data may offer some insight here.  Figure \ref{fig:tomo} shows cuts through the 3-D reconstructions of density and velocity for times covering the period of the LOFAR observations.  Each cut is in a plane running through the Sun-Earth line, but tilted to align with the line of sight to 3C147 (at an ecliptic latitude of 26$^{\circ}$).  The velocity reconstructions show a clear region of high velocity propagating across the 3C147 line of sight for the entire period of the LOFAR observations, suggesting that the higher, off-radial velocities seen in the short-baseline CCFs around 01:00\,UT may be the more-accurate ones (at least in terms of speed).  Indeed a high speed co-rotating region passed the Earth with a maximum velocity of over 700\,km\,s$^{-1}$ on 8  September 2017 at $\sim$12\,UT, and this is seen persisting to the north along the LoS to 3C147 in the 3-D reconstructions and so could be the source of these high velocities in the LOFAR data.  However, there is little sign of the CME material crossing the 3C147 line of sight (indicated by the black arrows in Figure \ref{fig:tomo}) in the density reconstructions.  The only possible indication is a tongue of slightly enhanced density gradually extending across the 3C147 line of sight throughout the period.  This appears more to follow behind the high-velocity region than be a direct part of it, which would be consistent with the LOFAR results, if any part of the high-velocity region is due to the CME material.  However, these reconstructions are necessarily low-resolution; for the September 2017 time period, only ISEE data are available and there were no LOFAR observations other than the ones presented here to feed into full reconstructions.  As a transit instrument, ISEE can only observe each radio source once a day, thus taking 24 hours to perform a single scan across the inner heliosphere.  For example, ISEE did observe 3C147, but only at around 21:00\,UT on 11 and 12 September and so would have seen no indication of the CME in these measurements \citep[][]{Iwaietal:2022}.  Other observations may have similarly missed information, leading to the density modelled by the tomography for this region being lower than it might otherwise be.  Thus, significant detail of the CME structure is likely missing from these constructions, providing a further illustration of how limited data may miss crucial information in some fast events.  \citet{Jacksonetal:2020} and \citet{Jacksonetal:2022} discuss this issue in more detail, and demonstrate how the reconstructions are improved through the incorporation of additional data sources.  

Information gleaned from both the tomographic reconstructions and the SUSANOO-CME mode suggests that the scattering seen in the LOFAR observations up to around 04:00\,UT or so is predominantly from slow solar wind close to the point of closest approach of the line of sight to the Sun, itself very close to the Earth at this elongation angle.  The high velocity seen coming to a peak two hours ahead of the peak in LOFAR g-level, and in the tomographic reconstructions ahead of the tongue of density crossing the 3C147 line of sight, may be associated with a shock front ahead of the main bulk of CME material.  A study by \citet{Manoharan:2010} of a CME from December 2006 using observations of IPS from the Ooty radio telescope in India revealed a similar structure: An initial rise in g-level was followed a few hours later by a much sharper, short-duration, increase, with a period of enhanced velocity peaking ahead of the peak in g-level.  This was interpreted as a shock front producing enhanced turbulence and velocity ahead of the CME itself, adding credence to the same interpretation being relevant here. The approaching CME also appears to push the background solar wind ahead and/or to the side of it in an off-radial direction (as noted for a different event by \citet{Breenetal:2008}), as seen in the lower plot in Figure \ref{fig:velocities}, where substantial off-radial deviations can be seen after 03:00\,UT.


The spatial correlation methods described in Section \ref{subsec:spatialcorrelation} offer a useful visualisation of the IPS data which can only be accomplished using an instrument with many observing stations and baselines covering hundreds of kilometres.  The necessity to correctly account for radio-source structure, particularly in single-site observations, when applying the weak scattering model of equation \ref{eqn:pspec} is obvious from Figure \ref{fig:source}.  This particularly affects estimates of the axial ratio of the density fluctuations (equation \ref{eqn:q}), which can be strongly biased by the source structure.  The amount of bias naturally depends on the solar wind conditions prevalent in any given observation, the radio source itself and it's orientation with respect to that of the solar wind density irregularities.  For the observations presented here an axial ratio of $\sim$1.3 may be estimated from the zero-lag spatial correlation prior to the arrival of the CME (Figures \ref{fig:zerolag} and \ref{fig:source}), and this is what may be obtained in a fit to a single-site observation, whereas the spatial correlation at a time-lag of 1.73\,s (Figure \ref{fig:spatialtimelags}) shows that the axial ratio is likely around 1.  Mitigation strategies for single-site observations of IPS include, obviously, using more-accurate source models where they exist, and making use of the fact that any orientation of source structure with respect to the radial direction is a slowly-varying phenomenon as the source approaches and recedes from the Sun and is itself constant year-to-year.  For the former strategy, the work of \citet{Chhetrietal:2018} may be extremely helpful in obtaining source structures of a large number of commonly-used IPS sources.

The cross-correlation analyses used here are much less susceptible to the effects of source structure: In particular, the non-zero spatial correlation images of Figure \ref{fig:spatialtimelags} separate quite neatly the moving structure of the scintillation due to the CME and/or background solar wind from static structure due to the radio source itself or noise in the data.  This allows direct estimates to be made of the density fluctuation structure of both the CME and the background solar wind and its motion across the line of sight.  The ability to do this for any observation is, however, limited by the available baselines.  These become heavily foreshortened in one direction when the radio source is at low elevations, limiting the spatial coverage which can be observed, and sparse baseline coverage leads to distortions in the images (e.g. Figure \ref{fig:spatialtimelags}d), although these are typically obvious.  For the observation in Figure \ref{fig:spatialtimelags} it can be seen that the CME structure is clearly elongated along the direction of a magnetic field strongly rotated away from the radial direction.  The background solar wind structure, by contrast, appears almost circular, suggesting that the density fluctuations are predominantly isotropic at this elongation from the Sun.

Since baseline coverage is best for the shorter baselines ($<\sim$500\,km), the background solar wind appears isotropic, and the effects of the CME are clearly dominant after around 05:00\,UT; the spatial correlations at zero time-lag can be used to assess the orientation of the CME density fluctuations and therefore the alignment of the CME magnetic field (Figures \ref{fig:spatialdirections} and \ref{fig:cmeangles}).  This shows strong rotation through the course of the observations, up to $\sim$80$^\circ$ off-radial.  This technique could therefore offer a different remote-sensing method to obtain some information on the interplanetary magnetic field wherever such observations of IPS are made.  There are naturally caveats to this: The orientation is plane-of-sky and the result of a line-of-sight integration; furthermore, it is not possible to determine magnetic-field {\it direction} by this method, only alignment and the rotation of this with time.  Validation of these results via MHD modelling, whilst strongly desired, would be an involved process warranting its own study and is therefore not performed here.  Plans are being developed to compare IPS results such as these with space weather models such as EUHFORIA for the purpose of assessing if IPS, as a source of fixed observations, can be used to help validate EUHFORIA model runs and/or prune ensembles.  Some initial studies combining IPS and MHD models have already been completed by \citet{Gonzietal:2021} where ISEE IPS and the UCSD tomography were used to investigate the impact of inner-heliospheric boundary conditions in MHD models on solar-wind predictions at Earth.  Such comparisons would also employ similar techniques to the comparison between pulsar observations and IPS tomography detailed in \citet{Tiburzietal:2022}, and between pulsar observations and EUHFORIA detailed in \citet{Shaifullahetal:2022}.

\section{Conclusions}

The techniques demonstrated here represent a significant advance in terms of visualising the various components that make up the overall IPS intensity pattern and thereby allowing initial parameters for a full application of the weak-scattering model to be estimated more accurately.  This includes the separation of radio-source structure effects, the density fluctuation scales of both a background solar wind and a CME, magnetic-field rotation associated with the CME, and velocity including direction for both solar wind and CME.

The background, slow, solar wind is found to deviate significantly from the radial direction due to the approach and passage of the CME material, while the CME itself continues to propagate in a mostly radial direction.  This suggests the CME was pushing the slow solar wind ahead and around the CME strongly to one side.

This event was the result of a merger between the ultra-fast CME of 10 September 2017 and slower CME material from eruptions a few days earlier.  A denser patch of material is seen from the SUSANOO-CME model to cross the LOFAR lines of sight, with some less dense material on its leading edge and behind it, and this appears reflected in the overall rise of scintillation g-level and a larger spike between $\sim$07:00 and 08:00\,UT (Figure \ref{fig:glevel}).

Velocity is seen to peak about two hours ahead of the peak in LOFAR g-level which may be associated with a shock front propagating ahead of the CME itself.  Tomographic reconstructions, although low in resolution, also appear to show this, with a region of high velocity seen propagating just head of a tongue of enhanced density crossing the 3C147 line of sight (Figure \ref{fig:tomo}).

Spatial correlations reveal rotation of the CME magnetic field to almost perpendicular to the radial direction, thus demonstrating a possible new method to obtain information on the magnetic-field orientation in the inner heliosphere.  Although it should be clearly noted that exact direction (i.e. North or South) cannot be established via this method at this time.

Since these observations were carried out, two spacecraft missions have been launched to take close-up in-situ measurements of the Sun, corona, and inner heliosphere - Parker Solar Probe \citep[][]{Foxetal:2016} and Solar Orbiter \citep[][]{Mulleretal:2020}.  The presence of these ground-breaking spacecraft at and inside of the locations typically observed using IPS offers the possibility to directly compare velocity and density measurements made using IPS with those taken in-situ much closer to the Sun.  LOFAR has been taking observations during periods when these spacecraft are taking measurements, meaning that the techniques demonstrated here can be applied to new data and combined with new in-situ information.

To conclude, this paper demonstrates the variety of solar wind and CME information which can be extracted from observations of IPS taken with an advanced instrument like LOFAR.  However, as a radio observatory for which observing time must be applied for and awarded based on scientific merit, it is not an instrument which currently observes the inner heliosphere full time.  The LOFAR4SpaceWeather (LOFAR4SW - \url{http://lofar4sw.eu}, see also the paper by \citet{Carleyetal:2020}) project is a design study which has undertaken a four-year investigation into the upgrades necessary to allow LOFAR to run space weather monitoring observations in parallel with radio astronomy operations, thus representing a first step towards enabling such monitoring full time on LOFAR.

\section{Acknowledgments}
This paper is based on data obtained with the International LOFAR Telescope (ILT) under project code DDT8\_006.  LOFAR \citep{vanHaarlemetal:2013} is the Low Frequency Array designed and constructed by ASTRON.  It has observing, data processing, and data storage facilities in several countries, that are owned by various parties (each with their own funding sources), and that are collectively operated by the ILT foundation under a joint scientific policy.  The ILT resources have benefitted from the following recent major funding sources: CNRS-INSU, Observatoire de Paris and Université d'Orléans, France; BMBF, MIWF-NRW, MPG, Germany; Science Foundation Ireland (SFI), Department of Business, Enterprise and Innovation (DBEI), Ireland; NWO, The Netherlands; The Science and Technology Facilities Council, UK; Ministry of Science and Higher Education, Poland.  Two of us (RAF and MMB) were partially supported by the LOFAR4SW project, funded by the European Community’s Horizon 2020 Programme H2020 INFRADEV-2017-1 under grant agreement 777442.  MMB was also supported in part by the STFC In-House Research grant to the Space Physics and Operations Division at UKRI STFC RAL Space.  KI was supported by MEXT/JSPS KAKENHI Grant Number 21H04517.  IPS observations of ISEE were made under the solar wind program of the Institute for Space-Earth Environmental Research, Nagoya University.  BVJ acknowledges funding from NASA contracts 80NSSC17K0684, 80NSSC21K0029, and AFOSR contract FA9550-19-1-0356 to the University of California, San Diego.

\bibliographystyle{model5-names}
\biboptions{authoryear}
\bibliography{refs}

\end{document}